# The Phase Diagram of High-Tc Cuprates

E. C. Marino[*]

*Instituto de Física, Universidade Federal do Rio de Janeiro,
C.P. 68528, Rio de Janeiro, RJ, 21941-972, Brazil.*

(Dated: August 25, 2024)

The detailed structure of the $T \times doping$ phase diagram of hole doped High-Tc superconducting cuprates is investigated from the perspective of a recently proposed comprehensive theory for these materials. Our theory is compared to Anderson's RVB theory for High-$T_c$ cuprates and it is demonstrated that the ground-eigenstate of our theory's Hamiltonian is a RVB-like state. From the Hamiltonian we derive the thermodynamic potential, as a function of convenient order parameters, through a method whose stability is carefully demonstrated. From this, we obtain analytic expressions for the transition lines delimiting the Néel phase, $T_N(x)$, the Spin-Glass phase, $T_g(x)$ and the CDW Charge Ordered phase, $T_{co}(x)$. These results, along with the previously derived expressions for the transition lines delimiting the Superconducting phase, $T_C(x)$, the Pseudogap and Strange Metal phases, $T^*(x)$ and the Fermi Liquid phase, $T_{FL}(x)$ are in excellent agreement with the experimental data from the cuprates. Our study reveals, in particular, the complementary role, which is played by the mechanisms responsible for the onset of superconducting (SC) and Spin-Glass (SG) phases in these materials. The absence of SG phases in electron doped High-Tc superconducting cuprates, strongly suggests a different SC mechanism should operate in those materials.

## 1) INTRODUCTION

Since the discovery of High-Tc superconductivity (HTSC) in cuprates, in 1986 [1], a large amount of theoretical effort has been made, in order to achieve the understanding, not only of the mechanism responsible for superconductivity (SC) in these materials, but also for the rich variety of different phases, which are observed in them. Despite of that, the theoretical results that had been obtained so far, were not able to reproduce, even partially, the vast amount of physical properties that were experimentally measured in High-Tc cuprates.

In a recent series of published studies, [2–6], we proposed a comprehensive microscopic theory for hole-doped cuprates, which, on the contrary, has demonstrated excellent agreement with the available experimental output.

The results we extracted so far from our theory provide an explanation for the superconducting (SC), pseudogap (PG), strange metal (SM) and Fermi liquid (FL) regimes of the SC cuprates. Each of these regimes is characterized by a pair of order parameters $(\Delta, M)$, which correspond, respectively, to the ground-state expectation values of Cooper pair ($\Phi$) and exciton ($\chi$) creation operators: $(\Delta = \langle\Phi\rangle, M = \langle\chi\rangle)$ [2]. We have $(\Delta \neq 0, M = 0)$ in the SC regime, $(\Delta = 0, M \neq 0)$ in the PG regime, $(\Delta = 0, M = 0)$ in the SM and FL regimes. These two are distinguished by the value of the chemical potential: $\mu \neq 0$ in the former and $\mu = 0$ in the latter [2–6].

Our theory has provided precise descriptions of a large number of physical observables, which agree very well with the set of experimental data, that emerges from measurements performed in several of these materials. We derived, in particular, analytical expressions, which successfully reproduce the results of experiments measuring physical observables such as: the superconducting (SC) and pseudogap (PG) transition temperatures as a function of the stoichiometric doping parameter $x$, namely, $Tc(x)$ and $T^*(x)$; the SC transition temperature as a function of the number of $CuO_2$ planes per primitive unit cell, N, namely $Tc(x, N)$; the SC transition temperature as a function of an applied external pressure, $Tc(x, P)$; the resistivity $\rho(T,x)$ in the non-SC phases, exhibiting, in particular, the linear-in-T behavior, observed in the SM phase and the $BT^2$ behavior in the FL phase (here we calculated the $B$ coefficient, as well, and it is in excellent agreement with the experimental result [3]); the resistivity in the presence of a constant magnetic field $H$, namely, $\rho(H,T,x)$, from which we derived the $H$ to $H^2$ crossover observed in the magnetoresistivity experiments, respectively in the strong and weak field regimes [4].

Using our theory we were able to determine the spectral density as a function of the PG order parameter $M$, namely, $n(\omega, M)$. Our theoretical result reproduces the measured depletion of states near the Fermi level, which characterizes the PG phase, wherever $M \neq 0$ [5]. This has confirmed that $M$, the ground-state expectation value of the exciton creation operator, is a good order parameter for the PG regime.

Besides that, the theory also makes a number of predictions. Among these we have: that the PG transition temperature $T^*(x)$ is not affected by an applied external pressure [2]. Then, our theory predicts that the excitons (electron-hole pairs) associated to the PG order parameter $M$ have spin 1. We also predict [3]) that the slope of the resistivity linear-in-$T$ dependence in the SM phase is proportional to the PG transition temperature $T^*(x)$.

In view of the above predictions, it follows that our theory is easily testable. As a matter of fact, actually, it has successfully met all the tests to which it has been submitted so far [2–6].

The above mentioned results were obtained for a variety of hole-doped cuprate materials including LSCO as well as the Hg, Bi and Tl families of cuprates. It turns out, however, that any theory supposed to describe the High-Tc cuprates must, forcefully, describe as well the full phase diagram of these materials. So far, we have used our theory to obtain a successful description of the SC, PG, SM and FL phases of the cuprates [2–6]. We have so far, consequently, an incomplete description of the HTSC cuprates.

In the present study, we are going to complete the above described picture by exploring the remaining phases of the $T \times x$ phase diagram of hole-doped cuprates to which our theory had not yet been applied. These cover a substantial region of the phase diagram, namely, the Néel, Stripes, Spin-Glass and CDW charge order phases.

The results we obtain include analytical expressions for the transition lines: the Néel temperature, $T_N(x)$; the Spin-Glass transition temperature $T_g(x)$; the CDW charge-ordering transition temperature, $T_{CO}(x)$. Such results, are presented below, together with the previously obtained [2–6]: Superconducting transition temperature, $T_C(x)$; Pseudogap transition temperature, $T^*(x)$; Fermi Liquid transition temperature, $T_{FL}(x)$. They are in excellent agreement with the experimental data for several High-Tc compounds, and ultimately allow for the obtainment of an unprecedentedly complete and accurate description of the phase diagram of hole doped cuprates.

Hole-doped High-Tc cuprates interestingly exhibit a phase diagram where a Spin-Glass (SG) phase recurrently appears in between the Superconducting (SC) and Néel phases. Being magnetically ordered insulators that, upon doping, become Spin-Glasses and eventually, upon further doping, superconductors, at sufficiently low temperatures, it is natural to inquire about the existence of an interplay between the SG and SC phases in these fascinating materials.

Indeed, the study we present here reveals the existence of a deep relation between the mechanism responsible for the Spin-Glass formation and the mechanism of superconductivity in hole-doped cuprates.

More specifically, we are going to demonstrate, in the present study, that the essential magnetic interaction that couples the spins of the itinerant doped holes, with the localized spins of the copper ions, is responsible both for the SG and SC phases in hole-doped cuprates. In the first case it randomly modifies the original AF Heisenberg coupling $J_{AF}$ either to $J_{AF}+J_0$ or to $J_{AF}-J_0$, according to whether the doped hole goes into the $p_x$ or $p_y$ oxygen orbitals. This creates the frustration that produces the SG phase. In the second case, the same relative $\pm$ sign, which produced the spin-glass, now makes nearest neighbor holes, in which one of the holes always belong to a $p_x$ and the other to a $p_y$ oxygen orbitals, to have opposite sign magnetic interactions with the closest localized spin. This leads to an effective *attractive* interaction between nearest neighbor hole pairs and produces a SC ground state, which is essentially the RVB state proposed by Anderson [7–13] for doped cuprates. In our theory, however, we explicitly show that an RVB-like ground-state is an eigenstate of the Hamiltonian proposed to describe the system.

We call the readers' attention to the fact that electron-doped cuprates do not exhibit SG phases [17]. This observation strongly suggests that a different mechanism of superconductivity operates in electron-doped cuprates. We are presently investigating what should be the mechanism for SC in electron-doped High-Tc cuprates.

The Kondo-like magnetic interaction that exists between localized and itinerant spins, produces an effective attractive interaction between the latter, when they belong to nearest neighbor sites of the oxygen lattice [2–6]. The on-site Coulomb repulsion between holes, conversely, was shown to be responsible for the onset of the PG phase [2–6], the condensation of excitons associated to a nonzero PG order parameter in this phase being directly responsible for the depletion of states around the Fermi level [5].

The studies that have led to the present theory for High-Tc SC in cuprates [2–6] are preceded, long time ago, by a series of publications which considered systems with Dirac electrons and explored the duality between SC phases and exciton condensates in such systems[18–23].

The article is organized in such a way that, in order to make it self-contained and reader-friendly, some published material was included, always with complete references. The purpose of this was to facilitate the readership and understanding of the arguments.

In Sect. 2, we present a concise review of the most relevant features of our theory and, in subsection 2.4, demonstrate that our Hamiltonian's ground-state is RVB-like. In Sect. 3, we apply our theory in the obtainment of a precise description of the phase diagram $T$ vs. $x$ of the cuprates, which includes detailed studies of the Néel, Spin Glass and CDW charge ordered phases. These, when combined with our previous results for $Tc$, $T^*$ and $T_{FL}$, produces a complete description of this phase diagram, with an accuracy with no precedents. In Sect. 4 we include a detailed comparison of our theory with Anderson's RVB theory, identifying what is missing in the latter, in order to become a predictive theory. We explain, in particular, why Anderson's theory, albeit starting from an inspired and correct conjecture about the nature of the ground-state, fails in reproducing the abundant experimental data, which are available for the cuprates. A discussion of several important issues raised by this study is presented in Sect. 5 and the main conclusions, in Sect. 6. We also include three Appendices: A) discussing the stability of the expansion about the ground state expectation values $\langle \Phi \rangle$ and $\langle \chi \rangle$; B) performing the operation leading to the spin-glass phase; C) investigating the specific case of YBCO according to our theory.

## 2) THEORY

The formulation of a theory for the High-Tc cuprates involves five basic steps, which are described below. The first step is the identification of the fundamental elementary microscopic constituents of the system, namely, the most convenient degrees of freedom to represent the system. This is a far-from-obvious step, and turns out to be an important ingredient for the eventual success of the theory. Then, the next step is the description of the basic interactions among the fundamental constituents. This is done, by means of a suitable Hamiltonian. This evidently depends on the choice made in the previous step. The third step is the obtainment of the eigenstates and eigenvalues thereof, which are going to describe the underlying physical properties of the system. The fourth step consists in using the energy eigenvalues for the derivation of the thermodynamic potentials that will contain all information about the system. The fifth step will be to extract the relevant information about the physical observables, from the thermodynamic potentials, as a function of $T, x, \mu$, which will be finally compared to the experimental data resulting from measurements.

Some examples are: a) the value of the order parameters that minimize these potentials, which will determine the different phases and the transition lines among them; b) the resistivity, which expresses the response of the system's electric current to an applied electric field, coupled to the thermodynamic potential; the magneto-resistivity, which is the previous response in the presence of an additional magnetic field; c) the variation of any of the previous observables as a function of an external hydrostatic pressure, which is introduced through the thermodynamic potential; d) the same for the dependence on the compounds number of $CuO_2$ planes per unit cell and so on.

### 2.1) The Basic Constituents and Their Interactions

It is a well-known fact that the cuprates are organized in a stack of $N = 1, 2, 3, ...$ $CuO_2$ planar structures, where all the relevant physical processes effectively occur. These planar systems possess electrons located in the $3d_{x^2-y^2}$ orbitals of copper and in the $p_x$ and $p_y$ orbitals of oxygen. The copper ions form a square lattice of spacing $a$, while the oxygen form a bipartite square lattice with spacing $a/\sqrt{2}$. The three intertwined square lattices form the 2d structure depicted in Fig. 1 [2, 5, 6].

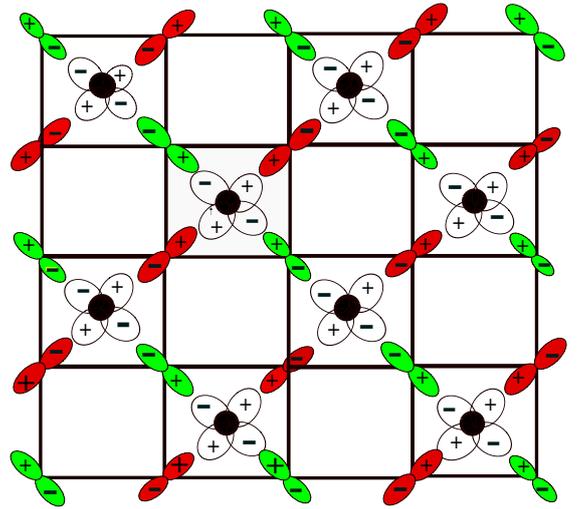

FIG. 1: The $CuO_2$ planar structure containing the localized copper-spins (black circles: d-orbitals) and the itinerant oxygen holes (red and green p-orbitals on sites $\mathbf{R}$ and $\mathbf{R} + \mathbf{d}$, respectively, of the $A, B$ oxygen sub-lattices) [2, 5, 6].

In a first approach, we may describe the system, exclusively in terms of the electrons belonging to these three types of orbitals and interacting through the repulsive electric interaction contained in the Hubbard terms of the Hamiltonian. This choice will lead to a description of the cuprates in terms of the Three Bands Hubbard Model (3BHM)[15, 16]. This purely electronic model has proven to be rather complicated, hence simplified versions thereof have been sought. We describe below two of the most relevant simplified versions of the 3BHM, namely, the t-J Model and the Spin-Fermion Model. Finally, we introduce our starting model: the Spin-Fermion-Hubbard model.

#### 2.1.1) The t-J Model

The t-J Model is obtained by applying the Schrieffer-Wolff (SW) transformation [24–27] to the Hubbard Hamiltonian and subsequently projecting the doubly occupied states out of the physical Hilbert space. This is a unitary canonical transformation, which was used to obtain the Kondo Hamiltonian, out of the Anderson impurity model [24]. It is also closely related to the Foldy-Wouthuysen transformation [28], which was used to obtain the Pauli-Schrödinger Hamiltonian with relativistic corrections, out of the Dirac Hamiltonian, describing relativistic spin 1/2 particles.

The t-J Model describes fermionic degrees of freedom (electrons or holes) with charge and spin that live on the sites of a simple square lattice having lattice spacing $a$. Perhaps the main draw-back of the t-J Model is the fact that the single square lattice where it is defined and where its relevant degrees of freedom move, does not show the detailed structure of the three intertwined lattices, where the real degrees of freedom actually live. As a consequence, the details of the spatial arrangements of the original $CuO_2$ planar structure are lost when we choose the t-J Model description of the system.

At half-filling, when we have exactly one fermion per site, the model reduces to a Heisenberg Hamiltonian, describing the spin-spin interactions among nearest neighbor fermions.

### 2.1.2) The Spin-Fermion Model

We now come to the Spin-Fermion-(SF) Model, which results from another simplification of the 3BHM. Perhaps the most important feature of this model is the fact that it preserves the spatial identity of the elementary constituents that live in each of the three intertwined square lattices that form the $CuO_2$ planes.

At zero doping, the fully occupied $p_x$ and $p_y$ orbitals of the $O^{--}$ ions forming the bipartite square lattice are completely inert. The $Cu^{++}$ active degrees of freedom are single electrons belonging to the $d_{x^2-y^2}$ orbitals of copper. Due to the very strong $U_{dd} \simeq 8.5eV$ electric repulsion between two electrons (or holes) in a copper atom, we can describe the copper degrees of freedom as localized spins $\mathbf{S}_i$ [29], forming a square lattice of spacing $a$. This, together with the oxygen ions, form the $CuO_2$ planar structure found in the cuprates.

There is clear experimental evidence [31, 32] that the electrons in the $Cu^{++}$ ions are localized states on the sites of a square lattice, which therefore we can associate to a primary operator located at site $i$, namely $\mathbf{S}_i$. It is also a well established fact that in hole-doped cuprates, the doped holes go into the otherwise completely filled $p_x$ and $p_y$ oxygen orbitals [31, 32].

If we choose the localized copper spins, $\mathbf{S}_i$, and the itinerant holes doped into the oxygen $p_x$ and $p_y$ orbitals, associated to the creation operators: $\psi^\dagger_{A,\sigma}$ and $\psi^\dagger_{B,\sigma}$ as the fundamental microscopic constituents of the cuprates, we are led to the simplified version of the 3BHM known as the Spin-Fermion Model [30]. This contains three terms:

$$H_{SF} = H_0 + H_{AF} + H_K \tag{1}$$

where $H_0$ represents the kinetic energy of the itinerant oxygen holes that move along the bipartite oxygen lattice. $H_{AF}$, the antiferromagnetic exchange interaction among the localized copper spins and $H_K$, the Kondo-like magnetic interaction between the itinerant oxygen spins (holes) and the localized copper spins.

The limit where there are no holes at all, but only localized copper spins, corresponds to the half-filling regime of the t-J Model, when that model describes a system with precisely one fermion per site. In this regime both the t-J Model and the SF Model reduce exactly to a Heisenberg antiferromagnet. This situation would describe the undoped parent compounds of the High-Tc cuprates.

As we dope holes into the system, according to the SF picture, they will go into the oxygen ions and become itinerant on a bipartite square lattice, while the number of localized copper spins will remain the same. Conversely, according to the t-J picture, as we dope holes into the system, the Hamiltonian will describe the out-of-half-filling regime of an all-itinerant fermion system.

### 2.1.3) Interactions: The Spin-Fermion-Hubbard Model on a Dimerized Lattice

**Hamiltonian**

The formulation of our theory is based on the Hamiltonian $H_{SFH} = H_{SF} + H_U$, where the first term is given by (1) and $H_U$ represents the on site Coulomb repulsion among holes. These four terms comprise what is called the Spin-Fermion-Hubbard (SFH) model [2–6], which is obtained from the Spin-Fermion (SF) model [30] by the inclusion of the electric repulsion term $H_U$.

$$\begin{aligned} H_0 &= -t_p \sum_{\mathbf{R},\mathbf{d}_i} \sum_\sigma \psi^\dagger_{A,\sigma}(\mathbf{R})\psi_{B,\sigma}(\mathbf{R}+\mathbf{d}_i) + hc \\ H_U &= U_p \sum_\mathbf{R} n^A_\uparrow n^A_\downarrow + U_p \sum_{\mathbf{R}+\mathbf{d}_i} n^B_\uparrow n^B_\downarrow \\ H_{AF} &= J_{AF} \sum_{\langle IJ \rangle} \mathbf{S}_I \cdot \mathbf{S}_J \\ H_K &= J_K \sum_I \mathbf{S}_I \cdot \left[ \sum_{\mathbf{R} \in I} \eta_A \eta_C\, \mathcal{S}_A + \sum_{\mathbf{R}+\mathbf{d} \in I} \eta_B \eta'_C\, \mathcal{S}_B \right] \end{aligned} \tag{2}$$

where $\psi^\dagger_{A,\sigma}(\mathbf{R})$, $\psi^\dagger_{B,\sigma}(\mathbf{R}+\mathbf{d})$ are the hole creation operators, with spin $\sigma=\uparrow,\downarrow$, on sites $\mathbf{R}$ and $\mathbf{R}+\mathbf{d}$, respectively, of the $A,B$ oxygen sub-lattices. $n^{A(B)}_\sigma = \psi^\dagger_{A(B),\sigma}\psi_{A(B),\sigma}$ are the hole number operators for the sublattices $A,B$. Finally, $\mathbf{S}_I$ are the spin operators of the localized $Cu^{++}$ ions.

In the previous expressions

$$\mathcal{S}_{A,B} = \frac{1}{2}\psi^\dagger_{(A,B)\alpha} \vec\sigma_{\alpha\beta} \psi_{(A,B)\beta} \tag{3}$$

are the spin operator of holes belonging to the $p_x,p_y$ oxygen orbitals, which are associated, respectively, to the $A$ and $B$ sub-lattices. The sign factors $\eta_A, \eta_B = \pm 1$ originate in the overlap integrals involving the atomic orbitals and are determined by the sign of the half-portion of the $p_x$ and $p_y$ oxygen orbitals that overlaps with the one-fourth portion of the $d_{x^2-y^2}$ copper orbitals, whose sign we denote by either $\eta_C = \pm 1$ or $\eta'_C = \pm 1$.

Because of the strong electric repulsion on the oxygen atoms [3], $U_{pp} \simeq 5.5 eV$, it turns out that the direct Oxygen-Oxygen jump of holes is strongly suppressed. Nevertheless, they escape from the cage through the mutual coupling of the $p_x$ and $p_y$ oxygen orbitals with the copper atom d-orbital. This indirect jump between neighbor $p_x$ and $p_y$ oxygen orbitals, via the d-orbital of copper is facilitated by the small size of the $pd$ electric repulsion, namely, $U_{pd} \simeq 0.897 eV$ [3].

**Lattice**

The peculiarities of the lattice structure hosting these microscopic constituents play a crucial role in the mechanism that produces not only the SC but also the normal phases of high-Tc cuprates.

Notice the organization of the signs of the $p_x$ and $p_y$ oxygen orbitals in a dimerized form:(- - + + - - + + - - + +), instead of a linear form: (- + - + - + - + - + - ) [5, 6]. These two configurations of signs of the holes' oxygen orbitals, would be degenerate in the absence of the localized copper spins, however the interaction thereof with the oxygen holes, energetically favors the dimerized form [5, 6].

This dimerization is similar to the Peierls mechanism that generates a gap in polyacetylene [34, 40] and also to the Yukawa mechanism that generates a mass to the quarks and leptons in the Standard Model [40]. In the high-Tc cuprates, the dimerization generates a SC gap as we shall see below.

In order to describe the doping process, we add to the above Hamiltonian the chemical potential term $-\mu \mathcal{N}$, where $\mu$ is the chemical potential of the holes and $\mathcal{N}$, the hole number operator:

$$\mathcal{N} = \sum_\sigma \left( \psi^\dagger_{A,\sigma} \psi_{A,\sigma} + \psi^\dagger_{B,\sigma} \psi_{B,\sigma} \right) \tag{4}$$

### 2.1.4 The Grand-Partition Function and the Effective Holes' Interaction

The Grand-Partition function corresponding to the Hamiltonian (2) is ($\beta = 1/k_B T$)

$$Z = \text{Tr}_\psi \left\{ e^{-\beta[H_0[\psi] + H_U[\psi] - \mu \mathcal{N}]} Z_{\mathbf{S}_I}[\psi] \right\} \qquad Z_{\mathbf{S}_I}[\psi] = \text{Tr}_{\mathbf{S}_I} \left\{ e^{-\beta[H_{AF}[\mathbf{S}_I] + H_K[\mathbf{S}_I, \psi]]} \right\} \tag{5}$$

The trace over the itinerant degrees of freedom (holes) can be accomplished by means of a functional integral over the fermion fields, which we assemble in the form of a Nambu fermion field [2–6], that is associated to the doped holes:

$$\Psi_a = \begin{pmatrix} \psi_{A,\uparrow,a} \\ \psi_{B,\uparrow,a} \\ \psi^\dagger_{A,\downarrow,a} \\ \psi^\dagger_{B,\downarrow,a} \end{pmatrix}, \tag{6}$$

where $a = 1, ..., N$ is the number of planes per primitive unit cell.

The trace over the localized degrees of freedom, conversely, can be performed by means of a bosonic functional integral over the ferromagnetic ($\mathbf{L}$) and antiferromagnetic ($\mathbf{n}$) fluctuation fields [40] of the localized copper spins [2–6], namely,

$$\text{Tr}_{\mathbf{S}_I} = \int D\mathbf{n} D\mathbf{L} \delta(|\mathbf{n}|^2 - 1)$$
$$\text{Tr}_\psi = \int D\Psi D\Psi^\dagger \tag{7}$$

Applying to $Z_{\mathbf{S}_I}[\psi]$, we obtain

$$Z_{\mathbf{S}_I}[\psi] = \text{Tr}_{\mathbf{S}_I} e^{-\beta H[\mathbf{S}_I, \psi]} =$$
$$\int D\mathbf{n} D\mathbf{L} \delta(|\mathbf{n}|^2 - 1) \exp \left\{ -\int_0^\beta d\tau \left[ -J_{AF} s^2 \sum_{<IJ>} \mathbf{n}_I \cdot \mathbf{n}_J + J_K s \sum_I \sum_{i \in I} (-1)^I \mathbf{n}_I \cdot [\eta_A \eta_C \; \mathcal{S}_A + \eta_B \eta'_C \; \mathcal{S}_B]_{i \in I} \right. \right.$$
$$\left. \left. 4 J_{AF} s^2 a^2 \sum_I |\mathbf{L}_I|^2 + \sum_I \sum_{i \in I} \mathbf{L}_I \cdot [J_K s [\eta_A \eta_C \; \mathcal{S}_A + \eta_B \eta'_C \; \mathcal{S}_B]_{i \in I} + i s \mathbf{n}_I \times \partial_\tau \mathbf{n}_I] \right] \right\} \tag{8}$$

Integrating over **L**, the ferromagnetic fluctuations field of the localized spins, we arrive at

$$Z_{\mathbf{S}_I}[\psi] = \exp\{-\beta H_1[\psi]\}\tilde{Z}_{NLSM} \tag{9}$$

where

$$H_1[\psi] = \frac{J_K^2}{8J_{AF}}\eta_A\eta_B\eta_C\eta_C' \sum_{\mathbf{R},\mathbf{R}+\mathbf{d}_i} \left[\psi^\dagger_{A\uparrow}(\mathbf{R})\psi^\dagger_{B\downarrow}(\mathbf{R}+\mathbf{d}_i) + \psi^\dagger_{B\uparrow}(\mathbf{R}+\mathbf{d}_i)\psi^\dagger_{A\downarrow}(\mathbf{R})\right]\left[\psi_{B\downarrow}(\mathbf{R}+\mathbf{d}_i)\psi_{A\uparrow}(\mathbf{R}) + \psi_{A\downarrow}(\mathbf{R})\psi_{B\uparrow}(\mathbf{R}+\mathbf{d}_i)\right] \tag{10}$$

and

$$\tilde{Z}_{NLSM} = \int D\mathbf{n}\,\delta(|\mathbf{n}|^2-1)\exp\left\{-\int_0^\beta d\tau\left[-J_{AF}s^2\sum_{<IJ>}\mathbf{n}_I\cdot\mathbf{n}_J + \sum_{<I>}+\frac{1}{c^2}\partial_\tau\mathbf{n}_I\cdot\partial_\tau\mathbf{n}_I\right.\right.$$
$$\left.\left.+J_K s\sum_I\sum_{i\in I}(-1)^I\mathbf{n}_I\cdot[\eta_A\eta_C\,\mathcal{S}_A + \eta_B\eta_C'\,\mathcal{S}_B]_{i\in I}\right]\right\} \tag{11}$$

is the partition function of the Nonlinear Sigma Model [2–6] augmented by the Kondo-like term, which describes the interaction between the itinerant spins and the antiferromagnetic fluctuations of the localized spins.

From Fig. 1, we see that for nearest neighbors belonging to different sub-lattices $A, B$ we will always have

$$\eta_A\eta_C\eta_B\eta_C' = -1. \tag{12}$$

Consequently the effective interaction between holes contains the term $H_1[\psi]$, which is always attractive, and originates from the magnetic mutual interaction between neighboring holes and localized copper spins.

The Grand Partition functional, then, can be expressed as

$$Z = \int D\Psi D\Psi^\dagger \left\{e^{-\beta[H_0[\psi]+H_U[\psi]-\mu\mathcal{N}]}Z_{\mathbf{S}_I}[\psi]\right\} \tag{13}$$

Now, making a 2nd. order perturbative expansion in $t_p$ on $H_0 + H_U$, in (5), we obtain [2]

$$Z = \int D\Psi D\Psi^\dagger \left\{e^{-\beta[H_0[\psi]+H_1[\psi]+H_2[\psi]-\mu\mathcal{N}]}\tilde{Z}_{NLSM}\right\} \quad;\quad Z = Z_{Holes}\tilde{Z}_{NLSM} \tag{14}$$

where, $H_1[\psi]$ is given by (10) while $H_2[\psi]$, by

$$H_2[\psi] = -\frac{2t_p^2}{U_p}\sum_{\mathbf{R},\mathbf{d}_i}\left[\psi^\dagger_{A\uparrow}(\mathbf{R})\psi_{B\uparrow}(\mathbf{R}+\mathbf{d}_i) + \psi_{A\downarrow}(\mathbf{R})\psi^\dagger_{B\downarrow}(\mathbf{R}+\mathbf{d}_i)\right]\left[\psi^\dagger_{B\uparrow}(\mathbf{R}+\mathbf{d}_i)\psi_{A\uparrow}(\mathbf{R}) + \psi^\dagger_{B\downarrow}(\mathbf{R}+\mathbf{d}_i)\psi_{A\downarrow}(\mathbf{R})\right], \tag{15}$$

The integration over **L** leads to the effective interaction responsible for the Cooper pair formation in cuprates. The second order perturbative expansion in $t_p$, conversely, leads to the interaction responsible for the formation of spin one excitons, which are bound-states of a pair of neighboring electron and hole, which belong to different oxygen sublattices. Excitons condense in the Pseudogap phase, where $M \neq 0$ and thereby generates all the Pseudogap phenomena, including the depletion of states near the Fermi level [2, 5].

The coupling parameters are, respectively,

$$g_S = \frac{J_K^2}{8J_{AF}} \quad g_P = \frac{2t_p^2}{U_p}. \tag{16}$$

We have shown [2, 5, 6] that the values obtained by inserting the constants $J_K, J_{AF}, t, U_p$ in the expressions above, are in excellent agreement with experiments.

## 2.2) The Superconducting and Pseudogap Order Parameters

The effective interaction among the doped holes can be summarized, according to (14), by the Hamiltonian

$$H_{eff}[\psi] = H_0[\psi] + H_1[\psi] + H_2[\psi]. \tag{17}$$

Applying a Hubbard-Stratonovitch transformation to the quartic fermion interactions in the Grand-Partition function, (14) we can rewrite our effective Hamiltonian in terms of the Hubbard-Stratonovitch fields $\Phi$ and $\chi$, as [2]

$$\begin{aligned} H_{eff}[\psi] = & -t_p \sum_{\mathbf{R},\mathbf{d}_i} \sum_\sigma \psi^\dagger_{A,\sigma}(\mathbf{R})\psi_{B,\sigma}(\mathbf{R}+\mathbf{d}_i) + hc \\ & + \frac{1}{g_S} \sum_{\mathbf{R},\mathbf{d}_i} \Phi^\dagger(\mathbf{R},\mathbf{d}_i)\Phi(\mathbf{R},\mathbf{d}_i) + \sum_{\mathbf{R},\mathbf{d}_i} \Phi(\mathbf{R},\mathbf{d}_i)\left[\psi^\dagger_{A\uparrow}(\mathbf{R})\psi^\dagger_{B\downarrow}(\mathbf{R}+\mathbf{d}_i) + \psi^\dagger_{B\uparrow}(\mathbf{R}+\mathbf{d}_i)\psi^\dagger_{A\downarrow}(\mathbf{R})\right] + hc \\ & + \frac{1}{g_P} \sum_{\mathbf{R},\mathbf{d}_i \in \mathbf{R}} \chi^\dagger(\mathbf{R},\mathbf{d}_i)\chi(\mathbf{R},+\sum_{\mathbf{R},\mathbf{d}_i} \chi(\mathbf{R},\mathbf{d}_i)\left[\psi^\dagger_{A\uparrow}(\mathbf{R})\psi_{B\uparrow}(\mathbf{R}+\mathbf{d}_i) + \psi^\dagger_{A\downarrow}(\mathbf{R})\psi_{B\downarrow}(\mathbf{R}+\mathbf{d}_i)\right] + hc \end{aligned} \tag{18}$$

The Hubbard-Stratonovitch fields satisfy the field equations [2, 5]

$$\Phi^\dagger(\mathbf{R},\mathbf{d}_i) = g_S \left[\psi^\dagger_{A\uparrow}(\mathbf{R})\psi^\dagger_{B\downarrow}(\mathbf{R}+\mathbf{d}_i) + \psi^\dagger_{B\uparrow}(\mathbf{R}+\mathbf{d}_i)\psi^\dagger_{A\downarrow}(\mathbf{R})\right] \tag{19}$$

and

$$\chi^\dagger(\mathbf{R},\mathbf{d}_i) = g_P \left[\psi^\dagger_{A\uparrow}(\mathbf{R})\psi_{B\uparrow}(\mathbf{R}+\mathbf{d}_i) + \psi^\dagger_{A\downarrow}(\mathbf{R})\psi_{B\downarrow}(\mathbf{R}+\mathbf{d}_i)\right] \tag{20}$$

Observe that $\Phi$ is the creation operator of spin zero Cooper pairs of holes belonging to different sublattices, whereas $\chi$ is the creation operator of spin one excitons formed by electrons and holes belonging to different sublattices.

The ground-state expectation value of these operators, namely, $\langle\Phi\rangle = \Delta$ and $\langle\chi\rangle = M$ are, respectively, order parameters for the SC and PG states of the system, the former being a Cooper pair condensate while the latter consists in an exciton condensate. They are given by [2, 5].

$$\Delta(\mathbf{k}) = \Delta\left[\cos k_+ a' - \cos k_- a'\right] \tag{21}$$

and also that

$$M(\mathbf{k}) = M\left[\cos k_+ a' - \cos k_- a'\right] \tag{22}$$

where $k_\pm = \frac{k_x \pm k_y}{\sqrt{2}}$.

The SC and PG order parameters both have a d-wave symmetry, namely, change the sign under a 90° rotation, and have nodal lines along the $\pm \hat{x}$ and $\pm \hat{y}$ direction.

## 2.3) The Thermodynamic Potentials

The bridge between the theory and experimental measurements is made by the thermodynamic potentials. Let us explore then how to build these bridges. We have

$$Z = Z_{Holes}\tilde{Z}_{NLSM} \tag{23}$$

Then, expanding the Hubbard-Stratonovitch fields around their respective ground-state expectation values, namely

$$\Phi = \Delta + \phi \quad ; \quad \chi = M + \zeta, \tag{24}$$

we can write the effective Grand-Partition functional for the holes as

$$Z_{Holes} \simeq Z[\Delta, M, \mu] \tag{25}$$

In Appendix A we present a complete analysis of the stability of this expansion and thereby establish its validity.

Then, we introduce a thermodynamic potential $\Omega[\Delta, M, \mu]$, which is a functional of the SC and PG order parameters, $\Delta$ and $M$, given, respectively, by (21) and (22) and of the chemical potential.

$$Z[\Delta, M, \mu] = \exp\left\{-\beta \Omega[\Delta, M, \mu]\right\} \tag{26}$$

Also, for $\tilde{Z}_{NLSM}$, define

$$\tilde{Z}_{NLSM} = \exp\left\{-\beta \Omega[\sigma, m^2]\right\} \tag{27}$$

The full thermodynamic potential is, therefore

$$\Omega = \Omega[\Delta, M, \mu] + \Omega[\sigma, m^2] \tag{28}$$

We are going to determine $\Omega[\Delta, M, \mu]$ here and $\Omega[\sigma, m^2]$ in Section 3.3.

In order to obtain the potential $\Omega = \Omega[\Delta, M, \mu]$, we write

$$Z_{holes} = \int D\Psi D\Psi^\dagger \exp\left\{\int d^2 r \int_0^\beta d\tau \left[\frac{|\Delta|^2}{g_S} + \frac{|M|^2}{g_P} + N\mu d(x)\right] + \Psi^\dagger \left[i\partial_\tau + \mathcal{H}[\Delta, M] - \mu\right]\Psi\right\} \tag{29}$$

where

$$\mathcal{H} - \mu = \begin{pmatrix} -\mu & \epsilon + M & 0 & \Delta \\ \epsilon + M^* & -\mu & \Delta & 0 \\ 0 & \Delta^* & \mu & -\epsilon - M^* \\ \Delta^* & 0 & -\epsilon - M & \mu \end{pmatrix}. \tag{30}$$

The fermion integration yields a determinant of $\mathcal{H} - \mu$ that can be expressed in terms of the eigenvalues of this operator, namely,

$$\varepsilon_\pm = \pm\sqrt{|\Delta|^2 + \left(\sqrt{\epsilon^2 + |M|^2} \pm \mu\right)^2} \equiv \pm\sqrt{\Delta^2 + \beta_\pm^2}, \tag{31}$$

The thermodynamic potential, then, after performing the Matsubara summation over the frequencies associated to the $i\partial_\tau$ term of the action, is given by [2]

$$\Omega(\Delta, M, \mu) = \frac{|\Delta|^2}{g_S} + \frac{|M|^2}{g_P} + Nd(x)\mu - 2TN\left(\frac{a}{2\pi}\right)^2 \int d^2k \left[\ln\cosh\left(\frac{\varepsilon_+}{2T}\right) + \ln\cosh\left(\frac{\varepsilon_-}{2T}\right)\right]. \tag{32}$$

$g_s$ and $g_P$ are the coupling parameters (16) , which assume different values for each cuprate and $d(x)$ is a function (to be determined) of the stoichiometric doping parameter $x$. $\epsilon(k_x, k_y)$ is the tight-binding (kinetic) energy, namely, [2]

$$\epsilon(k_x, k_y) = -2t[\cos k_+ a' + \cos k_- a'], \tag{33}$$

where $k_\pm = \frac{k_x \pm k_y}{\sqrt{2}}$ and $a' = \frac{a}{\sqrt{2}}$, with $a'$ and $a$ being, respectively, the lattice parameters of the oxygen and copper ions lattices.

The $d_{x^2-y^2}$ symmetric SC and PG order parameters are given, respectively, by (21) and (22).

### 2.4) The Ground-State

I demonstrate here that the RVB-like state, depicted schematically in Fig. 2 is the ground eigenstate of the effective hole Hamiltonian.

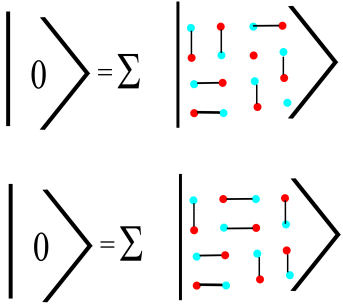

FIG. 2: The RVB-like ground-state in the SC phase. In the lower panel, we show the case when there is one hole per oxygen ion. In the upper panel, we show the corresponding ground-state for the case where we have less than one hole per oxygen ion. Here the red and cyan dots represent, respectively, the $p_x$ and $p_y$ completely filled (zero holes) orbitals of the oxygen atoms. Each link represents a Cooper pair formed by holes of opposite spins in nearest neighbor sites. The sum runs over all possible links, or, equivalently, degenerate states of crystalized links configurations (Spin Peierls).

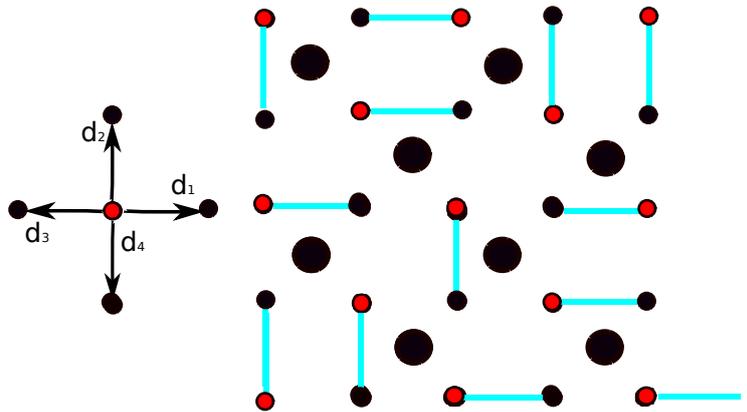

FIG. 3: One of the configurations contributing to the sum in (38) for the RVB-like ground-state when there is one hole per oxygen ion (see Fig. 1). Red and small black circles represent, respectively, oxygen ions with a single hole in the A and B sub-lattices. Large black circles represent the copper ions. The configuration is characterized by a distribution of connecting vectors belonging to the set $\{d_i[\mathbf{R}]\}$, where $d_i$, $i = 1, 2, 3, 4$ are represented by black arrows that correspond to the cyan bars (see figure). $\mathbf{R}$ runs over the sites of the A sub-lattice.

Let us start by defining the operators $C^\dagger(d_i[\mathbf{R}])$ and $E^\dagger(d_i[\mathbf{R}])$, which create, respectively, a Cooper pair formed by two neighbor holes with opposite spins and an exciton, formed by an electron and a neighbor hole, with the same spin. Both the Cooper pairs and the excitons, therefore, live on the links $l_i = d_i[\mathbf{R}], i = 1, ..., 4$, described in Fig. 3, which connect nearest neighbors belonging to the two oxygen sub-lattices. We call $L_\mathcal{N} = \{l_1, ..., l_{\mathcal{N}/2}\}$ a given configurations of links $l_i$ connecting the A and B sublattices in the presence of $\mathcal{N}$ doped holes. Let us call $\Omega_\mathcal{N}$ the set of all possible nearest neighbor link configurations for a given number of doped holes $\mathcal{N}$.

We now introduce the operators

$$C^\dagger(d_i[\mathbf{R}]) = C^\dagger(l_i) =$$
$$\left[\psi^\dagger_{A\uparrow}(\mathbf{R})\psi^\dagger_{B\downarrow}(\mathbf{R}+\mathbf{d}_i) + \psi^\dagger_{A\downarrow}(\mathbf{R})\psi^\dagger_{B\uparrow}(\mathbf{R}+\mathbf{d}_i)\right]$$
$$E^\dagger(d_i[\mathbf{R}]) = E^\dagger(l_i) = \sum_\sigma \psi^\dagger_{A\sigma}(\mathbf{R})\psi_{B\sigma}(\mathbf{R}+\mathbf{d}_i) \tag{34}$$

where the $\sigma$ sums run over $\sigma = \uparrow, \downarrow$. We can now write

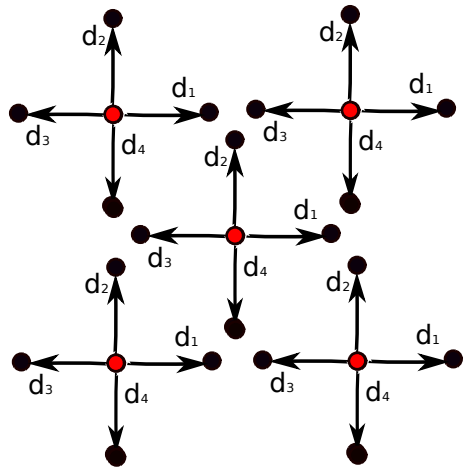

FIG. 4: Assembling (five) $A$-sites (red dots) with four appended links (black dots), each. Notice that the combined sum over $\mathbf{R}$, which runs over the sites of the A sub-lattice, plus the sum over the four link vectors attached to $\mathbf{R}$: $d_i[\mathbf{R}]$, with $i = 1, 2, 3, 4$, counts only once each of the links connecting the A and B sub-lattices.

the effective Hamiltonian (17) in the convenient form, in terms of the above operators:

$$H_0[\psi] = -t \sum_{\mathbf{R}} \sum_{\mathbf{d}_i \in \mathbf{R}} \left[ E(d_i[\mathbf{R}]) + E^\dagger(d_i[\mathbf{R}]) \right]$$

$$H_1[\psi] = -g_S \sum_{\mathbf{R}} \sum_{\mathbf{d}_i \in \mathbf{R}} C^\dagger(d_i[\mathbf{R}]) C(d_i[\mathbf{R}])$$

$$H_2[\psi] = -g_P \sum_{\mathbf{R}} \sum_{\mathbf{d}_i \in \mathbf{R}} E^\dagger(d_i[\mathbf{R}]) E(d_i[\mathbf{R}]) \quad (35)$$

Observe that the sums above run only once over all the horizontal and vertical links of the bipartite oxygen lattice.

Now, introducing the state with no holes $|0\rangle$, using the anti-commutation rules

$$\{\psi_{A\sigma}(\mathbf{R}), \psi_{A\sigma'}(\mathbf{R}')\} = \psi_{A\sigma}^\dagger(\mathbf{R}), \psi_{A\sigma'}^\dagger(\mathbf{R}')\} = 0$$

$$\{\psi_{B\sigma}(\mathbf{R}+\mathbf{d}_i), \psi_{B\sigma'}(\mathbf{R}+\mathbf{d}_i)'\} = \{\psi_{B\sigma}^\dagger(\mathbf{R}+\mathbf{d}_i), \psi_{B\sigma'}^\dagger(\mathbf{R}+\mathbf{d}_i)'\} = 0$$

$$\{\psi_{A\sigma}(\mathbf{R}), \psi_{A\sigma'}^\dagger(\mathbf{R}')\} = \delta_{\sigma,\sigma'} \delta_{\mathbf{R},\mathbf{R}'} \quad \{\psi_{B\sigma}(\mathbf{R}+\mathbf{d}_i), \psi_{B\sigma'}^\dagger(\mathbf{R}+\mathbf{d}_i)'\} = \delta_{\sigma,\sigma'} \delta_{\mathbf{R}+\mathbf{d}_i,(\mathbf{R}+\mathbf{d}_i)'} \quad (36)$$

and the fact that hole destruction operators acting on $|0\rangle$ give zero, it is not difficult to see that

$$E_i C_i^\dagger |0\rangle = \sum_\sigma \left[ \psi_{B\sigma}^\dagger(\mathbf{R}+\mathbf{d}_i) \psi_{A\sigma}(\mathbf{R}) + \psi_{A\sigma}^\dagger(\mathbf{R}) \psi_{B\sigma}(\mathbf{R}+\mathbf{d}_i) \right] C_i^\dagger |0\rangle = 0 \quad (37)$$

Also, noticing that the structure between square brackets in (34) corresponds to the cyan bar in Figs. 2, 3, and therefore belongs to one of the link configurations $L_\mathcal{N} \in \Omega_\mathcal{N}$, we introduce the state

$$|GS\rangle = \sum_{L_\mathcal{N} \in \Omega_\mathcal{N}} \left\{ \prod_{l_i \in L_\mathcal{N}} C^\dagger(l_i) \right\} |0\rangle \quad (38)$$

It follows that

$$H_1[\psi]|GS\rangle = -g_S \sum_{L_\mathcal{N} \in \Omega_\mathcal{N}} \left\{ \sum_{\mathbf{R}} \sum_{\mathbf{d}_i \in \mathbf{R}} C^\dagger(d_i[\mathbf{R}]) C(d_i[\mathbf{R}]) \right\} \left[ \prod_{l_j \in L_\mathcal{N}} C^\dagger(l_j) \right] |0\rangle \quad (39)$$

The non-vanishing terms of (39) are

$$H_1[\psi]|GS\rangle = -g_S \sum_{L_\mathcal{N} \in \Omega_\mathcal{N}} \left[ \prod_{l_j \in L_\mathcal{N}} C^\dagger(l_j) \right] |0\rangle \quad (40)$$

such that

$$H_1[\psi]|GS\rangle = -g_S \frac{\mathcal{N}}{2} |GS\rangle \quad (41)$$

The result (37) implies that $H_0[\psi]|GS\rangle = 0$ It is also straightforward to show that $H_2[\psi]|Gs\rangle = 0$. Then, using (41), we conclude that

$$H_{eff}[\psi]|GS\rangle = -g_S \frac{\mathcal{N}}{2} |GS\rangle \quad (42)$$

Using the completeness of the states obtained by acting with products of the operators $C_i^\dagger$, $E_i^\dagger$ and $\psi^\dagger$ on $|0\rangle$, it is not difficult to see that any other state that we may obtain from $|0\rangle$ will have more energy than $|GS\rangle$, which is, therefore, the ground-state.

Our effective Hamiltonian $H_{eff}$ is very similar to the Rokshar-Kivelson Hamiltonian, [12], which has been used [12] for describing the resonating Cooper pairs in the short-range RVB-state. We see that it can be derived from the underlying theory provided we make a convenient choice of the fundamental constituents of the cuprates and their mutual interactions.

### 3) THE PHASE DIAGRAM

#### 3.1) The Superconducting Phase: Tc(x)

The stationary condition is

$$\frac{\partial \Omega}{\partial \Delta} = 0 \; ; \frac{\partial \Omega}{\partial M} = 0 \; ; \frac{\partial \Omega}{\partial \mu} = 0 \quad (43)$$

Using the above conditions in the case $\Delta \neq 0, M = 0$

and then taking the limit $\Delta \to 0$, we find the transition temperature $T_c$ is given by [2]

$$\begin{cases} T_c(x) = \dfrac{\ln 2 \; T_{max}}{\ln 2 + \dfrac{\mu_C(x)}{2T_c(x)} - \dfrac{1}{2}\left(1 - e^{-\frac{\mu_C(x)}{T_c(x)}}\right)}, & x < x_0 \\ \\ T_c(x) = \dfrac{\ln 2 \; T_{max}}{\ln\left[1 + \exp\left[-\dfrac{\mu_C(x)}{T_c(x)}\right]\right]}, & x > x_0 \end{cases} \quad (44)$$

. The chemical potential, $\mu$, is a natural function of the amount of holes $y$ doped into the $CuO_2$ planes, namely, $\mu = \mu(y)$. $y$, by its turn, is a function of the stoichiometric doping parameter $x$, namely, $y = f(x)$. $x$ is the parameter that controls the doping [2] and the one, in terms of which, all the experimental consequences of doping are referred.

Knowledge of $x$, in general, does not imply knowledge of the chemical potential $\mu(y)$, because the function $y = f(x)$ is generally unknown. In order to circumvent this obstacle, we directly write $\mu = \mu(x)$, in terms of the stoichiometric doping parameter and immediately conclude that optimal doping occurs when the $x$ parameter has the value $x = x_0$, such that $\mu(x_0) = 0$. Then, it follows that the chemical potential along the curve $T_c(x)$, can be expressed as [2]

$$\mu_C(x) = 2\gamma(x_0 - x), \quad (45)$$

and $\gamma$ is a parameter which must be determined for each compound.

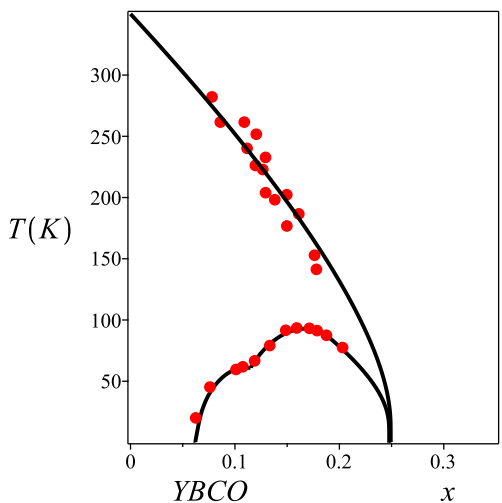

FIG. 5: The $YBCO$ $T \times x$ phase diagram (partial view). The continuous lines correspond to analytic expressions provided by our theory for the high-Tc cuprates (44) and (47) [2–6] (see Appendix C for details about YBCO). The experimental data are taken from [43].

$T_{max}$ is the transition temperature at optimal doping, $x_0$, which is given by

$$T_{max} = \frac{\Lambda \eta(Ng_S)}{2\ln 2} \quad (46)$$

where $\eta(Ng_S) = 1 - \frac{g_c}{Ng_S}$ and $g_S = \frac{J_K^2}{8J_{AF}}$.

$\Lambda = \frac{hv}{\xi} = 0.018$ eV is a characteristic energy scale associated to the coherence length $\xi$, $N$ is the number of $CuO_2$ planes per unit cell and $g_c$ lower threshold for the coupling parameter, namely $g_S > g_c = 0.30$ $eV$ [2–6].

Observe that for LSCO we use a symmetrized version of the equations to comply with the experimental observation that the SC dome is symmetrical for this compound.

### 3.2) The Pseudogap and Strange Metal Phases: T*(x)

In order to obtain the PG transition temperature $T^*(x)$, we use the stationary condition (43) for the case $\Delta = 0$ and $M \neq 0$, and subsequently take the limit $M \to 0$. This procedure leads to the following equation for $T^*(x)$[2]

$$T^*(x) = \frac{\frac{\Lambda \tilde{\eta}(g_P N)}{2}}{\ln\left[1 + \exp\left[-\frac{\tilde{\mu}(x)}{T^*(x)}\right]\right]}, \quad (47)$$

In the above equation [2]

$$\tilde{\mu}(x) = 2\tilde{\gamma}(\tilde{x}_0 - x), \quad (48)$$

is the chemical potential along the curve $T^*(x)$. $\tilde{x}_0$ is a parameter that will determine the point where $T^* \to 0$ and $\tilde{\gamma}$ must be determined for each compound [2, 5, 6].

In the Supplementary Material of [5], we present several figures depicting the curves $T_C(x)$ and $T^*(x)$, respectively taken from (44) and (47) in comparison to the corresponding measured values, for different cuprate compounds. The relevant parameters for each compound can also be found there and in [2].

### 3.3) The Néel Phase: $T_N(x)$

Let us rewrite the field associated with the antiferromagnetic fluctuations of the localized spins as $\mathbf{n} = (\pi_1, \pi_2, \sigma)$. Then, after integration over the ferromagnetic fluctuations of the localized spins, $\mathbf{L}$ and also over the transverse components, $\pi_1, \pi_2$, we find the effective action for the $\sigma$-component fields is given by

$$\Omega\left[\sigma, m^2\right] = \frac{1}{2} \int_0^{\hbar\beta} d\tau \int d^2r \left[\frac{1}{c^2}|\partial_\tau \sigma|^2 + |\nabla\sigma|^2 + m^2\left[\sigma^2 - \rho_0/3\right]\right] - 2\text{Tr} \ln\left[\frac{1}{c^2}\partial_\tau^2 + \nabla^2\right] \quad (49)$$

where $\rho_0$ is the spin-stiffness.

In order to obtain the Néel temperature, $T_N(x)$, we take the stationary conditions corresponding to the $\sigma$ and $m^2$ variables, namely,

$$\frac{\partial \Omega}{\partial \sigma} = 0 \; ; \frac{\partial \Omega}{\partial m^2} = 0 \quad (50)$$

obtaining, respectively, the equations

$$\sigma m^2 = 0$$
$$\sigma^2 = \frac{\rho_0}{3} - 2T \sum_{n=-\infty}^{\infty} \int \frac{d^2k}{(2\pi)^2} \frac{1}{\omega_n^2 + k^2 + m^2}$$
$$= \frac{\rho_0}{3} - \frac{1}{2\pi} \int_m^{\sqrt{\Lambda_a^2 + m^2}} dy \coth\left(\frac{y}{2T}\right) \quad (51)$$

Here $\Lambda_a = 2\pi\frac{\hbar c}{a}$ is a short-distance cutoff, which is naturally provided by the lattice parameter. A Néel phase is characterized by a nonzero sublattice magnetization: $\sigma \neq 0$. The first equation above, then implies $m = 0$. This, however, introduces a serious divergence in the integral in (51), which thereby would preclude a Néel phase in the strictly 2d system. This is a clear manifestation of the Mermin-Wagner theorem [36]. Considering, however, that the protagonist characters of our system, namely the doped holes, actually live in a 3d box of height $d$, the inter-plane distance, it follows that the $k$-integral in (51) covers a range $k > \kappa \equiv \frac{\hbar c}{d}$

$$\sigma^2 = \frac{\rho_0}{3} - \frac{1}{2\pi} \int_\kappa^{\Lambda_a} dy \coth\left(\frac{y}{2T}\right) = \frac{\rho_0}{3} - \frac{\Lambda_a}{2\pi} + \frac{\kappa}{2\pi} + \frac{T}{\pi} \ln\left(1 - e^{-\frac{\kappa}{T}}\right) \quad (52)$$

or, equivalently,

$$\sigma^2 = \frac{1}{\pi}\rho_s(T) \quad (53)$$

where

$$\rho_s(T) = \rho_s + T \ln\left(1 - e^{-\frac{\rho_s}{T}}\right) \quad (54)$$

Here $\rho_s = \rho_0/3$.

After we dope the system it is clear we shall have a doping dependent spin stiffness $\rho_s(x)$, whereupon we find the doping dependent Néel temperature, which corresponds to $\rho_s(T) = 0$, in (54), namely

$$T_N(x) = \frac{\rho_s(x)}{\left|\ln\left[1 - e^{-\rho_s(x)/T_N(x)}\right]\right|} \quad (55)$$

The threshold doping for the occurrence of the Néel phase, $x = x_{AF}$ is defined as $T_N(x_{AF}) = 0$. Consequently, we must have $\rho_s(x_{AF}) = 0$. Choosing the simplest parametrization for the spin stiffness, satisfying this condition, we write

$$\rho_s(x) = \rho_s\left(1 - \frac{x}{x_{AF}}\right). \quad (56)$$

and $\rho_s(0) = \rho_s$.

From (55) we obtain the Néel temperature at zero doping:

$$T_N(x = 0) = \frac{\rho_s}{\ln 2}. \quad (57)$$

### 3.3.1) Stripes

Here we must consider the possible presence of stripes in cuprates such as LSCO, for example. Without stripes we should have an isotropic system with the same spin stiffness along the x- and y-directions: $\rho_s^X(x) = \rho_s^Y(x)$, each of them given by (56). In the presence of stripes, conversely, we must have $\rho_s^X(x) \neq \rho_s^Y(x)$. In this case the effective spin stiffness would be given by [19, 35] the geometrical average of $\rho_s^X(x)$ and $\rho_s^Y(x)$.

Admitting that stripes in LSCO and other cuprates form only parallel to the y-axis, we shall assume that the process of doping only affects the spin stiffness in the

x-direction, hence we have

$$\rho_s^X(x) = \rho_s \left(1 - \frac{x}{x_{AF}}\right) \quad ; \quad \rho_s^Y(x) = \rho_s \quad (58)$$

$$\rho_s(x) = \sqrt{\rho_s^X(x)\rho_s^Y(x)} = \rho_s \left(1 - \frac{x}{x_{AF}}\right)^{1/2}$$

The following table contains the relevant parameters for the obtainment of the Néel temperature LSCO, Hg1201 and YbCO.

|        | $x_{AF}$ | $\rho_s$ (eV) | $\Lambda_a$ (eV) |
|--------|----------|---------------|-------------------|
| LSCO   | 0.020    | 0.0194        | 0.1843            |
| Hg1201 | 0.040    | 0.0318        | 0.1843            |
| YBCO   | 0.048    | 0.0258        | 0.1843            |

TABLE I: The parameters used for obtaining the $T_N(x)$ curves.

The Néel temperature for LSCO corresponding to these parameters and given by (55) is depicted as the solid line in Fig.6.(See [19] for the analogous results for Dirac fermions). The one corresponding to Hg1201, in Fig. 7 and the one for YBCO, in Fig. 8.

The sublattice magnetization at $T = 0$, as a function of doping, namely, $M(x) \equiv \sqrt{\rho_s(x)}$ is depicted in Fig.9, for LSCO, both in the presence of stripes or not. The experimental data, clearly indicates the presence of stripes in LSCO.

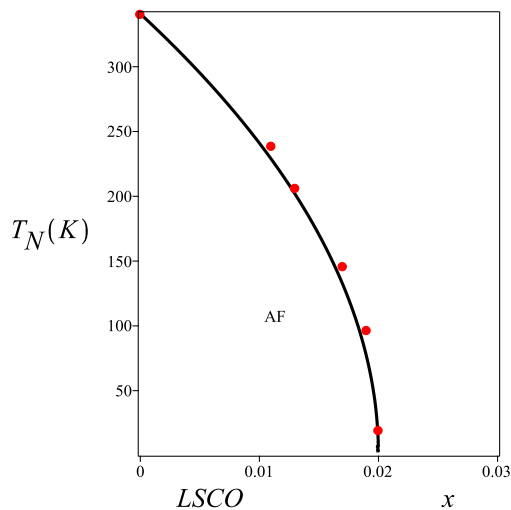

FIG. 6: $T \times x$ phase diagram of LSCO (partial view), showing the magnetically ordered Néel phase, delimited by the solid line, representing the Néel temperature $T_N$ predicted by our model, according to the analytical expression (55). The experimental data are taken from [41].

The sub-lattice magnetization at $T = 0$ is defined as $M(x) = \sqrt{\rho_s(x)}$, in terms of the spin stiffness. Hence, it can be expressed, respectively by

$$M(x) = M(0) \left(1 - \frac{x}{x_{AF}}\right)^{1/4} \quad (59)$$

and

$$M(x) = M(0) \left(1 - \frac{x}{x_{AF}}\right)^{1/2} \quad (60)$$

in the presence or absence of spin stripes. These two expressions are depicted in Fig.7 and comparison with experimental data clearly suggests the presence of stripes.

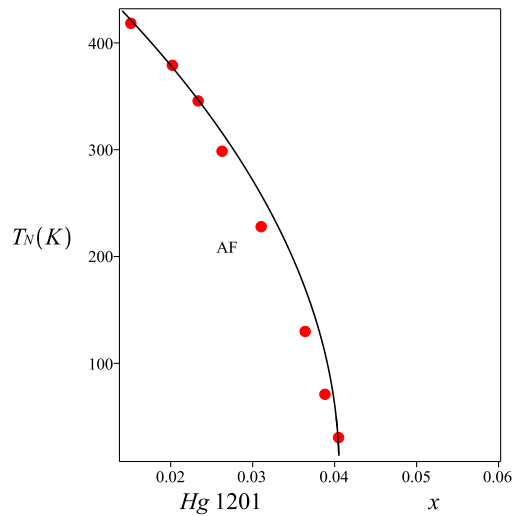

FIG. 7: $T \times x$ phase diagram of Hg1201 (partial view), showing the magnetically ordered Néel phase, delimited by the solid line, representing the Néel temperature $T_N$ predicted by our model, according to the analytical expression (55). The experimental data are taken from [37]

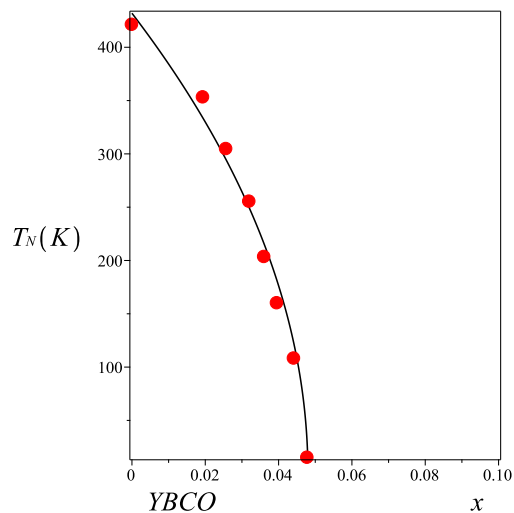

FIG. 8: $T \times x$ phase diagram of YBCO (partial view), showing the magnetically ordered Néel phase, delimited by the solid line, representing the Néel temperature $T_N$ predicted by our model, according to the analytical expression (55). The experimental data are taken from [43].

### 3.4) The Spin-Glass Phase: $T_g(x)$

Early attempts to describe the SG phase of High-Tc cuprates can be found in [33]. In the present study, however, it has become clear, for the first time, that the SG phase results from the (**n**) Antiferromagnetic fluctuations component of the magnetic Kondo-like interaction between holes and localized spins, whereas the SC phase results from the (**L**) Ferromagnetic component thereof, according to (7),(8),(11). The determination of the transition line $T_g(x)$, also had not been been achieved as yet.

The grand-partition functional can be written as $Z = Z_{holes}\tilde{Z}_{NLSM}$, where $\tilde{Z}_{NLSM}$ is given by (11) and $Z = Z_{holes}$, by (32)

In Appendix C, we do 2nd. order perturbation theory $J_K$ in (11), and thereby obtain an effective coupling $J_{AF}$ for the Heisenberg interaction among the localized spins whenever the doped holes do not go into the oxygen atoms connected to these spins. Conversely, we will have an effective localized spin coupling $J_{AF} \pm J_0$, where

$$J_0 = \frac{4J_K^2}{J_{AF}}, \tag{61}$$

and the plus or minus signs will hold, according to whether the doped hole goes into an oxygen atom belonging to sublattice $A$ or $B$. We immediately conclude we have a stochastic system with a partition functional given by

$$Z[J_{IJ}] = \int D\mathbf{n}\delta(|\mathbf{n}|^2 - 1) \exp\left\{-\int_0^\beta d\tau \left[s^2 \sum_{<IJ>} J_{IJ}\mathbf{n}_I \cdot \mathbf{n}_J + \sum_{<I>} \frac{1}{c^2}\partial_\tau \mathbf{n}_I \cdot \partial_\tau \mathbf{n}_I\right]\right\} \tag{62}$$

where the local couplings $J_{IJ}$ are random and given by

$$J_{IJ} = \begin{cases} J_{AF} \\ J_{AF} + J_0 \\ J_{AF} - J_0 \end{cases}$$

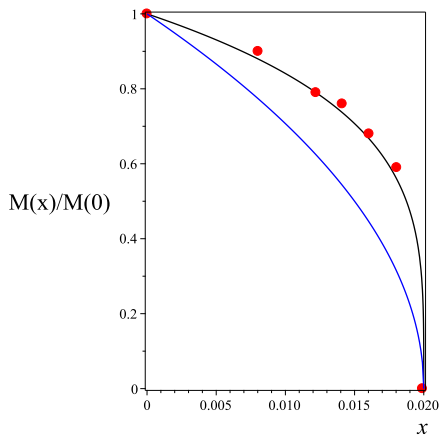

FIG. 9: Sub-lattice magnetization of LSCO. The black solid line corresponds to an effective spin stiffness given by (59), which was derived by assuming an anisotropic situation associated with the presence of stripes. The solid blue line corresponds to the isotropic situation, that would occur in the absence of stripes, where the spin stiffness would be given by (56). Experimental data are taken from [38] and clearly indicate the presence of stripes.

The associated probabilities are

$$\begin{cases} P[J_{AF}] = 1 - x \\ P[J_{AF} + J_0] = \frac{x}{2} \\ P[J_{AF} - J_0] = \frac{x}{2} \end{cases}$$

where $x$ is the doping parameter.

The average coupling and variance will be given, respectively, by

$$\langle J_{IJ} \rangle = J_{AF}$$
$$\Delta J_{IJ} = \sqrt{\langle J_{IJ}^2 \rangle - \langle J_{IJ} \rangle^2}$$
$$\Delta J_{IJ} = \sqrt{(1-x)[J_{AF}^2] + \frac{x}{2}[J_{AF}^2 + J_0^2 + 2J_{AF}J_0] + \frac{x}{2}[J_{AF}^2 + J_0^2 - 2J_{AF}J_0] - J_{AF}^2} = \sqrt{x}J_0 = \frac{4\sqrt{x}J_K^2}{J_{AF}} \tag{63}$$

The opposite nature of the magnetic interaction existing between the doped holes belonging to different sublattices and localized copper spins, namely Antiferromagnetic and Ferromagnetic, is responsible both for producing the effective attractive interaction between 1st neighbor holes, which leads to the superconducting phase in cuprates, as well as for the frustration that produces a spin-glass phase preceding the SC dome.

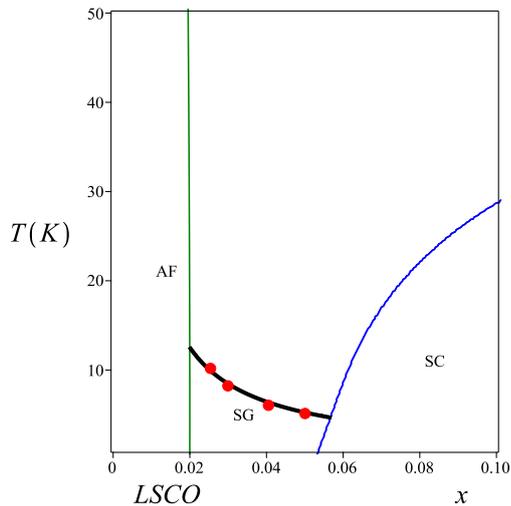

FIG. 10: The $LSCO$ $T \times x$ phase diagram (partial view). The continuous lines correspond to analytic expressions provided by our theory for the high-Tc cuprates for: a) the Néel transition line (green); b) for the Spin-Glass transition, line (black), expression (68) c) for the Superconducting transition line (blue) [2–6]. The experimental data for the Spin-Glass transition (red circles) are taken from [42].

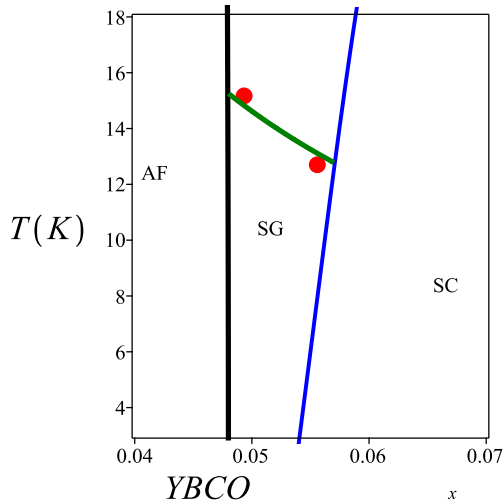

FIG. 11: The $YBCO$ $T \times x$ phase diagram (partial view). The continuous lines correspond to analytic expressions provided by our theory for the high-Tc cuprates for: a) the Néel transition line (green); b) for the Spin-Glass transition, line (black), expression (68) c) for the Superconducting transition line (blue) [2–6]. The experimental data for the Spin-Glass transition (red circles) are taken from [42].

Assuming a quenched situation, we use the so called replica method, in order to evaluate averages in the presence of frustration, namely

$$\langle \Omega \rangle = -k_B T \lim_{n \to 0} \left[ \frac{\langle Z^n[J_{IJ}] \rangle - 1}{n} \right] \quad (64)$$

Then, considering the possible values of the magnetic coupling and the respective probabilities corresponding to them, it follows that

$$\langle Z^n \rangle = \prod_{\alpha=1}^{n} \int D\mathbf{n}^\alpha \delta(|\mathbf{n}^\alpha|^2 - 1) \left\{ \frac{x}{2} \exp\left\{ -s^2 [J_{AF} + J_0] \int_0^\beta d\tau \sum_{<IJ>} \sum_{\alpha=1}^n \mathbf{n}_I^\alpha \cdot \mathbf{n}_J^\alpha \right\} \right.$$

$$+ \frac{x}{2} \exp\left\{ -s^2 [J_{AF} - J_0] \int_0^\beta d\tau \sum_{<IJ>} \sum_{\alpha=1}^n \mathbf{n}_I^\alpha \cdot \mathbf{n}_J^\alpha \right\}$$

$$\left. + (1-x) \exp\left\{ -s^2 [J_{AF}] \int_0^\beta d\tau \sum_{<IJ>} \sum_{\alpha=1}^n \mathbf{n}_I^\alpha \cdot \mathbf{n}_J^\alpha \right\} \exp\left\{ -\sum_{<I>} \sum_{\alpha=1}^n \frac{1}{c^2} \partial_\tau \mathbf{n}_I^\alpha \cdot \partial_\tau \mathbf{n}_I^\alpha \right\} \quad (65)$$

or, equivalently,

$$\langle Z^n \rangle = \prod_{\alpha=1}^{n} \int D\mathbf{n}^\alpha \delta(|\mathbf{n}^\alpha|^2 - 1) \exp\left\{ -\int_0^\beta d\tau \left[ s^2 J_{AF} \sum_{<IJ>} \sum_{\alpha=1}^n \mathbf{n}_I^\alpha \cdot \mathbf{n}_J^\alpha + \sum_{<I>} \sum_{\alpha=1}^n \frac{1}{c^2} \partial_\tau \mathbf{n}_I^\alpha \cdot \partial_\tau \mathbf{n}_I^\alpha \right] \right\} \times$$

$$\left[ 1 + x \left[ \cosh\left( \int_0^\beta d\tau s^2 J_0 \sum_{<IJ>} \sum_{\alpha=1}^n \mathbf{n}_I^\alpha \cdot \mathbf{n}_J^\alpha \right) - 1 \right] \right] \quad (66)$$

Considering that $x$ is a small parameter ($x < 0.05$ for LSCO), we can write

$$1 + x \left[ \cosh\left( \int_0^\beta d\tau s^2 J_0 \sum_{<IJ>} \sum_{\alpha=1}^n \mathbf{n}_I^\alpha \cdot \mathbf{n}_J^\alpha \right) - 1 \right] \simeq$$

$$\exp\left\{ \frac{s^4 x J_0^2}{2} \int_0^\beta d\tau \int_0^\beta d\tau' \sum_{<IJ>} \sum_{\alpha,\beta=1}^n \mathbf{n}_{a,I}^\alpha(\tau) \mathbf{n}_{b,I}^\beta(\tau') \mathbf{n}_{a,J}^\alpha(\tau) \mathbf{n}_{b,J}^\beta(\tau') \right\} \quad (67)$$

where $a, b = 1, 2, 3$ are the Cartesian components of $\mathbf{n}$.

The last term in (66) introduces a quartic interaction term in the NLSM. We can see that the stochastic NLSM becomes replaced by a non-stochastic but interacting spin system with a coupling parameter proportional to $(\Delta J)^2 = x J_0^2$.

The result above is identical to the one obtained from a previous analysis of the Heisenberg antiferromagnet with a random coupling on a square lattice [40, 46–48]. The results obtained in that previous study, consequently, apply here as well.

A particularly useful result was the obtainment of the transition line $T_g(x)$, delimiting the SG phase, which is characterized by infinite on-site time correlations that are associated to a nonzero Edwards-Anderson SG order parameter. This is given by [40, 46–48]

$$T_g = \frac{\pi [\rho_s - \rho_0(x)]}{\ln\left[\frac{\Lambda_a}{T_g}\right] - \frac{1}{2} \ln[1 + \varphi]} \quad (68)$$

Here the spin stiffness $\rho_s$ is a quantum critical point, beyond which there can be no SG phase at $T = 0$ ($\rho_0(x) \leq \rho_s$).

We can express this in terms of the inverse frustration parameter

$$\varphi = 3\pi \left( \frac{\langle J \rangle}{\Delta J} \right)^2 \quad (69)$$

in the form [40, 46–48]

$$\pi [\rho_s - \rho_0(x)] = \lambda \varphi^{3/4} \quad (70)$$

We use the above formulas for $T_g(x)$, in order to describe the onset of a SG in cuprates For LSCO, we have $\lambda = 0.0085$, $\Lambda_a = 0.1843$ and

$$\varphi = \left( \frac{0.0107}{x} \right).$$

The comparison of the results with the experimental data are excellent, as we can see from Figs. 8 and 9.

### 3.5) The CDW Charge Order Phase: Tco(x)

Let us investigate here the possibility of existence of a charge (CDW) ordered phase in the framework of our theory for the cuprates. For this purpose, we consider the thermodynamic potential, $\Omega(\Delta, M, \mu)$, given by (32). Then, consider the stationary condition

$$\frac{\delta \Omega}{\delta M(\mathbf{k})} = 0, \tag{71}$$

which implies,

$$2M(\mathbf{k}) \left\{ \frac{1}{g_P} - \frac{N}{yT} \frac{\sinh y}{\cosh y + \cosh \frac{\mu}{T}} \right\} = 0, \tag{72}$$

where

$$y = \frac{\sqrt{\epsilon(\mathbf{k})^2 + M(\mathbf{k})^2}}{k_B T} \tag{73}$$

where $\epsilon(\mathbf{k})^2 = V^2 (\cos k_1 a + \cos k_2 a)^2$ is the tight-binding energy eigenvalue and $M^2 (\cos k_1 a - \cos k_2 a)^2$ is the DDW Pseudogap order parameter [2].

The Fermi surface can be defined as the manifold for which the eigenvalues of $H - \mu N$ vanish. In this case, the points of reciprocal space belonging to the Fermi surface satisfy (73), for $y = y_0 \neq 0$ [2].

For $M \neq 0$, we obtain, from (72) it can be shown that it

$$\alpha(x, T) y = \tanh y \tag{74}$$

where

$$\alpha(x, T) = \frac{T}{N g_P} \frac{\left[ \cosh y + \cosh\left(\frac{\mu(x)}{T}\right) \right]}{\cosh y}. \tag{75}$$

For $\alpha > 1$, the only solution is $y_0 = 0$. Consequently, a Fermi surface does not form in this case. For $\alpha < 1$, on the other hand, a Fermi surface will appear.

The threshold for this to happen, therefore, is $y_0 = 0$. The boundary of the region of the phase diagram, where a Fermi surface starts to be seen (Fermi pockets) is, then given by

$$\frac{T_{co}}{N g_P} = \frac{1}{\left[ 1 + \cosh\left(\frac{\mu(x)}{T_{co}}\right) \right]}. \tag{76}$$

The function between brackets on the rhs is monotonically decreasing, hence we must have the maximum of $T_{CO}$ at $T_{CO;max} = T_{CO}(x_{CO;max})$, such that $\mu_{CO}(x = x_{CO;max}) = 0$. Then, the simplest parametrization for $\mu_{CO}(x)$ will be (see [2])

$$\mu_{CO}(x) = 2\gamma_{CO}(x_{CO;max} - x) \tag{77}$$

where $x_{CO;max}$ is the experimental value of the doping for which the Charge Ordering temperature is maximal and $\gamma_{CO}$ is the only parameter that must be adjusted to fit the experimental data.

Now, from (76), we have $T_{CO;max} = \frac{N g_P}{2}$

It follows, consequently, that the equation

$$T_{CO} = \frac{2 T_{CO,max}}{1 + \cosh\left(\frac{2\gamma_{CO}(x_{CO;max} - x)}{T_{CO}}\right)} \tag{78}$$

determines the boundary of the region of the phase diagram for which the system present CDW charge ordering.

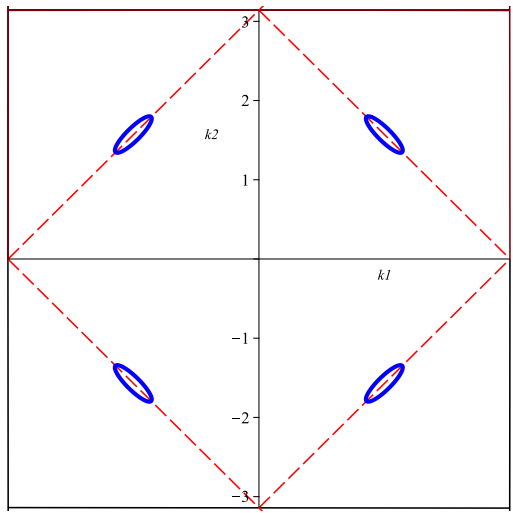

FIG. 12: The 1st Brillouin Zone of *LSCO*, showing the formation of a Fermi Surface (Elliptic Fermi Pockets, in blue).

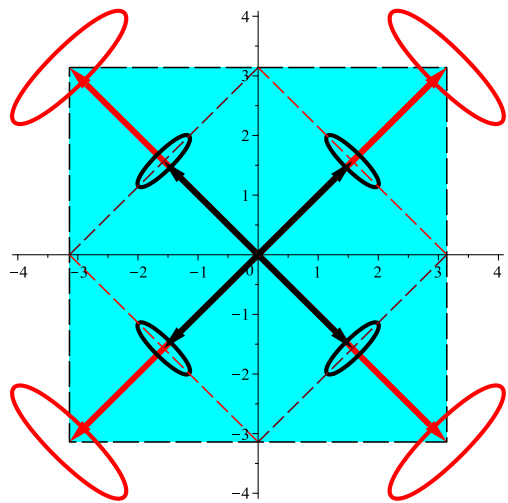

FIG. 13: The *LSCO* 1st Brillouin Zone, showing the doubling of the pocket Fermi vectors (in black), which are going to be the reciprocal lattice vectors where the CDW charge ordering occurs. Notice that these will be in the intersection of the 1st Brillouin Zone with the doubled ellipses (in red).

| | $T_{CO,max}$ (K) | $T_{CO,max}$ (eV) | $x_{CO,max}$ | $\gamma_{CO}$ (eV) |
|---|---|---|---|---|
| LSCO | 78.90 | 0.0068 | 0.114 | 0.092 |
| Hg1201 | 214.09 | 0.01845 | 0.088 | 0.3498 |
| YBCO | 151.13 | 0.01302 | 0.121 | 0.140 |

TABLE II: The parameters used for obtaining the $T_{co}(x)$ curves. Only $\gamma_{co}$ has been adjusted.

The following table contains the relevant parameters for LSCO, Hg1201 and YBCO.

We may determine the Fermi pockets explicitly, by expanding the functions, $\epsilon(\mathbf{k})^2$ and $M(\mathbf{k})^2$ around the points $(k_1, k_2) = (\pm\frac{\pi}{2a}, \pm\frac{\pi}{2a})$. These are four ellipses centered at such points, with semi-axes given, respectively, by (in $rlu$, reciprocal lattice units)

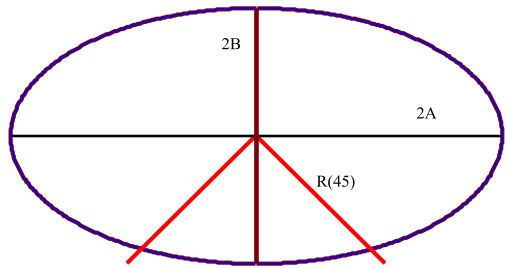

FIG. 14: The *LSCO* CDW charge ordering vectors, which correspond to the intersection of the quadrant of the 1st Brillouin Zone with the ellipse with doubled semi-axes.

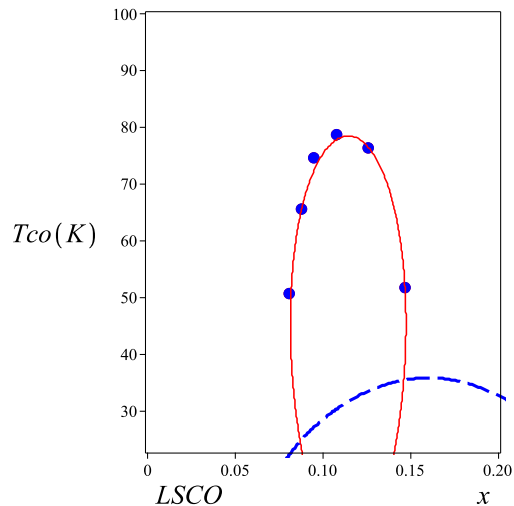

FIG. 15: $T_{co} \times x$ phase diagram of LSCO, showing the CDW charge ordered phase, delimited by the solid line, representing the $T_{co}$ temperature predicted by our theory, according to the analytical expression (78). The experimental data are taken from [44]. The dashed line represents our analytical expression for the SC dome.

$$A = \frac{y_0 k_B T}{2\pi\sqrt{2}V} \quad ; \quad B = \frac{y_0 k_B T}{2\pi\sqrt{2}\sqrt{(y_0 k_B T)^2 - V^2}}. \tag{79}$$

where $V = \hbar c/a$ and where we used the fact that, for $y_0 \neq 0$,

$$M^2 = (y_0 k_B T)^2 - V^2 \tag{80}$$

As it turns out, experimentally, the CDW charge order occurs in the same region of the phase diagram as the Fermi surface (pockets) formation. Hence we assume the usual situation in which charge Density Wave (CDW) order occurs, with a wave-vector $\mathbf{Q}_{CDW}$ that is twice the Fermi vectors, when the elliptical Fermi pockets start to form. We will be able, thereby, to successfully predict the CDW ordering wave-vector for a certain temperature and doping.

Indeed, starting from

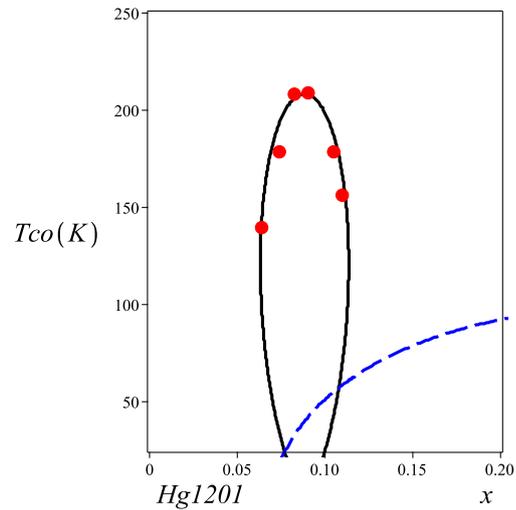

FIG. 16: $T_{co} \times x$ phase diagram of Hg1201, showing the CDW charge ordered phase, delimited by the solid line, representing the $T_{co}$ temperature predicted by our theory, obtained from the analytical expression (78). The experimental data are taken from [45]. The dashed line represents our analytical expression for the SC dome.

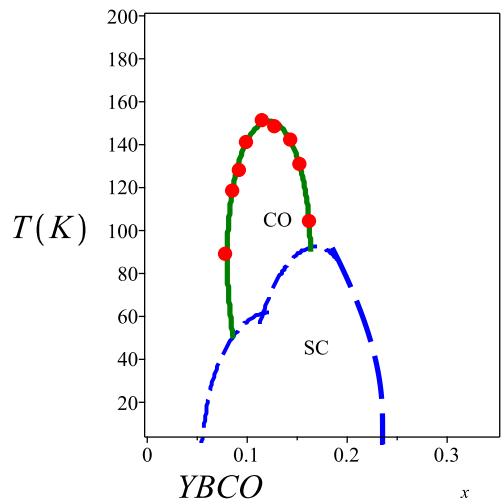

FIG. 17: $T_{co} \times x$ phase diagram of YBCO, showing the CDW charge ordered phase, delimited by the solid line, representing the $T_{co}$ temperature predicted by our theory, obtained from the analytical expression (78). The experimental data are taken from [45]. The dashed line represents our analytical expression for the SC dome.

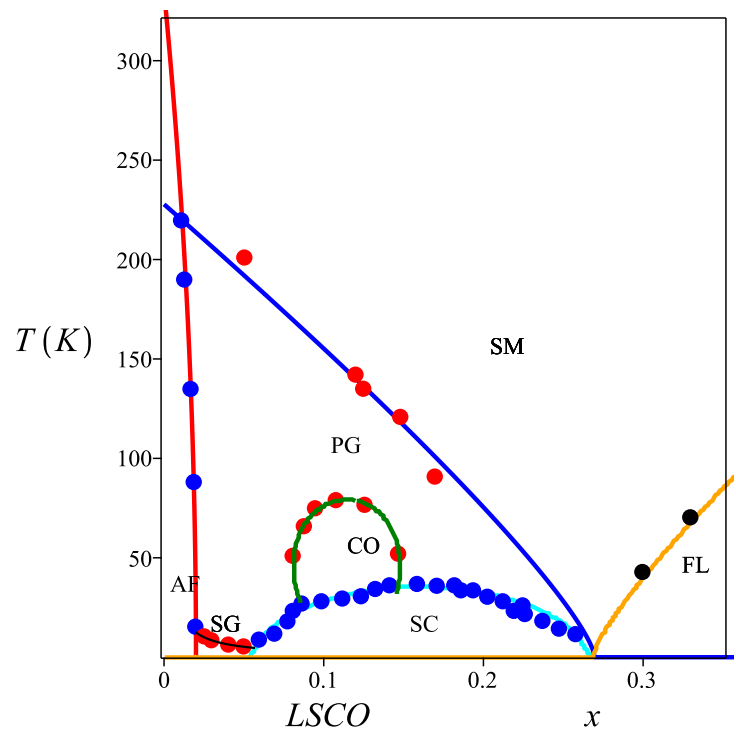

FIG. 18: The $LSCO$ $T \times x$ phase diagram. The continuous lines correspond to analytic expressions provided by our theory for the high-Tc cuprates for the transition temperatures: $T_N(x)$ (Néel-red),$T_g(x)$(Spin-Glass-black), $T_c(x)$ (SC transition-cyan),$T_{CO}$ (CDW-Charge Ordering-green) $T^*(x)$(PG transition-blue), $T_{FL}(x)$; $T_{FL}(x) = T^*(x - x_+)$ (FL transition-orange)[2–6].

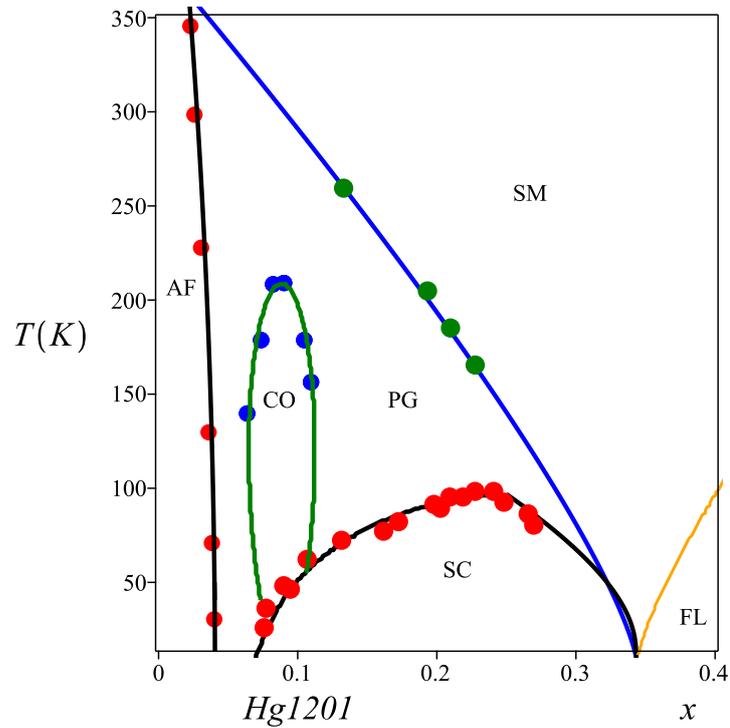

FIG. 19: The $Hg201$ $T \times x$ phase diagram. The continuous lines correspond to analytic expressions provided by our theory for the high-Tc cuprates for the transition temperatures: $T_{CO}$ (CDW-Charge Ordering-green), $Tc(x)$ (SC transition-black) $T^*(x)$(PG transition-blue), $T_{FL}(x)$(FL transition-orange)[2–6].

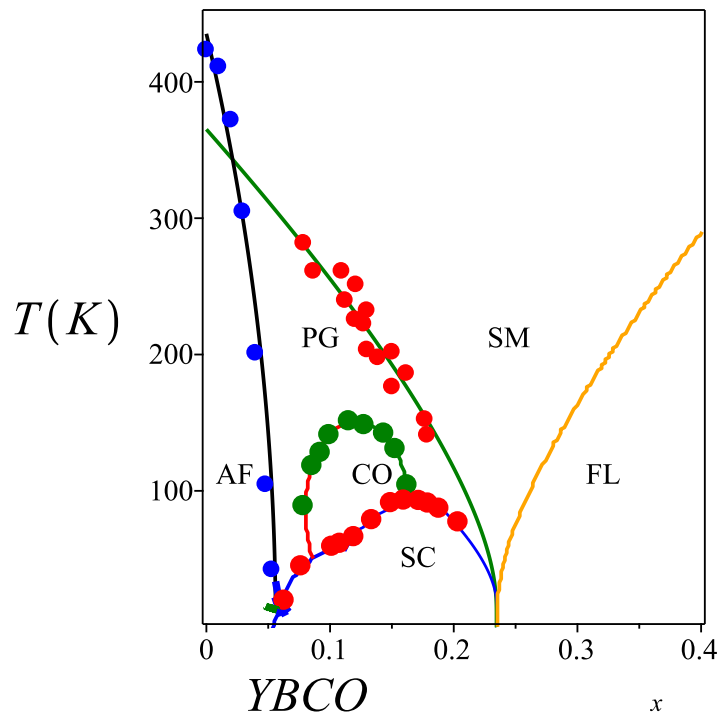

FIG. 20: The $YBCO$ $T \times x$ phase diagram. The continuous lines correspond to analytic expressions provided by our theory for the high-Tc cuprates for the transition temperatures: $T_{CO}$ (CDW-Charge Ordering-green), $Tc(x)$ (SC transition-black) $T^*(x)$(PG transition-blue), $T_{FL}(x)$(FL transition-orange)[2–6].

$$Q_{CDW} = 2\mathbf{q}_F \tag{81}$$

We conclude, therefore, that the CDW charge ordering vectors belong to an ellipse with semi-axes: $Q_A = 2A$ and $Q_B = 2B$. However, only the vectors belonging to the sector of the ellipse that is in the first Brillouin zone count. This is a region of the ellipse located within an angle $\theta \in [-45°, 45°]$, around the smallest semi-axis: $Q_B$. This immediately allows us to determine the range of magnitude of the CDW-ordering wave-vector, namely,

$$Q_B \leq \mathbf{q}_{CDW} \leq R(45°)$$
$$R(45°) = \frac{\sqrt{2} Q_A Q_B}{\sqrt{Q_A^2 + Q_B^2}}. \tag{82}$$

For LSCO at $T = 80K$, we have (in r.l.u)

$$Q_B = 2B = 0.232 \ ; \ \ R(45°) = \frac{2\sqrt{2}AB}{\sqrt{A^2+B^2}} = 0.318 \tag{83}$$

hence we find

$$0.232 \ r.l.u. \leq \mathbf{q}_{CDW} \leq 0.318 \ r.l.u., \tag{84}$$

which is in excellent agreement with the experimental data [49–53]

The function $T_{co}(x)$, expressed by (78) is depicted in Fig.13, Fig. 14 and Fig.15, respectively, for LSCO, Hg1201 and YBCO.

### 3.6) The Fermi Liquid Phase: $T_{FL}(x)$

The Fermi Liquid (FL) phase starts at the quantum critical point $\tilde{x}_0$, where $T^*(\tilde{x}_0) = 0$. The quantum critical phase associated to this point is delimited by $T^*(x)$ at the left and by $T^*(2\tilde{x}_0 - x)$ at the right of it. The right-hand curve delimits the FL phase, hence the FL transition temperature is given by $T_{FL}(x) = T^*(2\tilde{x}_0 - x)$.

The FL transition line is depicted (in orange) in Fig.16, Fig. 17 and Fig.18, respectively, for LSCO, Hg1201 and YBCO.

### 4) COMPARISON WITH THE RVB THEORY

The RVB theory, originally proposed by Anderson [8], similarly to Laughlin's theory for the Quantum Hall Effect [39], starts by hypothesizing the ground-state, rather than systematically obtaining it from a given Hamiltonian [14]. This hypothetical ground-state-vector of the system in the SC phase, proposed by Anderson, is the famous RVB state, which is very similar to the ground-state we found for our model and is depicted in Figs. 2,3. Yet, there are important differences that are listed below.

In our case the resonating dimers are pairs of holes with opposite spins belonging to neighbor sites (A,B) of the bipartite lattice of oxygen atoms, whereas in the Anderson RVB state the dimers belong to sites, (not necessarily nearest neighbors, according to [14]) of a generic square lattice, necessarily different from the oxygen bipartite lattice.

I have demonstrated, in preceding Subsections, that our proposed RVB-like state is the lowest energy eigenstate of the effective Hamiltonian $H_0 + H_1 + H_2$ we obtained for the holes after integrating out the localized degrees of freedom associated with the localized copper spins $\mathbf{S}_i$.

The situation where the RVB state exhibits a complete saturation of links, has different meanings in Anderson's theory and in ours. In the former, it corresponds to the half-filled case, where there is one hole per copper atom locked in a Heisenberg-type system. In our theory, conversely, it corresponds to a situation where there is precisely one hole per oxygen ion.

At this point I come back to the observation about the importance of a convenient choice of the relevant degrees of freedom. Our choice of localized copper spins plus itinerant oxygen holes to represent the system has produced, upon integration over localized degrees of freedom, an effective Hamiltonian for the holes which has a RVB-like ground-state.

This scenario is only possible because the relatively small value of $U_{pd} \simeq 0.9 eV$ allows the holes to become itinerant with assistance of the copper ions. Otherwise they would be locked, because of the high value of $U_{pp}$.

The knowledge of the actual Hamiltonian allows for the determination of the grand-partition function, whereby we obtain the thermodynamic potential, which is the bridge to make contact of the model with reality, namely the experimental data. Indeed, from the thermodynamic potential we can obtain the values of the order parameters. From these the phase diagram can be determined, as well as the transition curves which delimit them. Furthermore, the effect of an external electric or magnetic fields can be easily introduced in the thermodynamic potential, out of which the resistivity and magnetoresistivity can be obtained. Pressure effects on the phase diagram, can also be conveniently described through the thermodynamic potential [2–6].

Even though the knowledge of the correct ground-state alone, such as the RVB state in Anderson's theory, represents a great advance on the obtainment of a theory for the cuprates, the establishment of a comprehensive theory for these materials cannot be achieved without knowing the Hamiltonian of the system, which is preceded by a convenient choice of the relevant degrees of freedom. Concerning this fact, the path set forth by the Spin-Fermion-Hubbard Model seems to be the most appropriate.

## 5) DISCUSSION

1) The choice of the t-J Model to represent, in a simplified way, the SC cuprates involves a mixing of the physically localized, copper degrees of freedom, with the physically itinerant degrees of freedom of the oxygen ions. Mixing all of them in a common description does not seem to be the most convenient choice.

2) The model that served as the starting point for the development of our theory for High-Tc superconductivity in hole-doped cuprates, is the Spin-Fermion-Hubbard Model. Even though it is a well-known model, most of the researchers chose to follow the path that goes through to the t-J Model and their descendants.

3) An important ingredient of our theory is a subtle rearrangement of the signs (dimerization) of the oxygen $p_x$ and $p_y$ orbitals [5]: (+ - - + + - - +) instead of (+ - + - + - + -), which minimizes the energy of the system of oxygen holes, in the presence of the localized copper spins. This is crucial for the onset of a SC phase preceded by a SG phase. These two patterns would be degenerate for a non-interacting system, but, when coupled to a lattice of localized spins, becomes energetically favorable to the dimerized form, which produces both the attractive interaction among holes belonging to nearest neighbors and the stochastic frustration responsible for the onset of a spin-glass phase at low-doping, in between the AF and SC phases.

The effective interaction existing between any two nearest neighbor holes, belonging respectively to a $p_x$ and a $p_y$ oxygen orbitals acquires a negative sign, as a consequence of the above described dimerization, and becomes *attractive* [5, 6]. Without dimerization, the sign of (**??**) would be reversed and the RVB-like state would not be the ground-state.

The rearrangement of the oxygen p-orbitals is a dimerization [5], similar to what happens in the Peierls mechanism in polyacetylene. We conclude that the mechanisms responsible both for the existence of the SC and SG phases of the hole doped High-Tc cuprates depend crucially on the dimerization pattern of the oxygen sublattices, in the $CuO_2$-planes.

4) The magnetic interactions of the localized copper spins $\mathbf{S_i}$ can be decomposed into two components, which are mediated by $\mathbf{L}$ and $\mathbf{n}$, respectively, the ferromagnetic and antiferromagnetic fluctuations fields. When interacting with the magnetic moments of the oxygen doped holes, the former produces a hole-hole attraction that leads to Cooper pair formation and Superconductivity, while the latter produces a random increase/decrease on the effective $Cu - Cu$ exchange parameter that introduces Spin-Glass phase.

5) In the latter case, the effective magnetic interaction between an arbitrary pair of nearest neighbor localized copper spins acquires an effective exchange coupling that can have three possible values, namely, $[J_{AF}, J_{AF} + J_0, J_{AF} - J_0]$. The 1st one applies when there are no holes in the oxygen orbitals near the localized spins. The last two apply when there is one hole in such orbitals, and the $\pm$ signs occur according to whether the hole goes, with equal probability either into the $p_x$ and $p_y$ orbitals, thus creating the frustration responsible for the Spin-Glass behavior.

## 6) CONCLUSION

There are two avenues that start at the Three Bands Hubbard Model and aim at the complete description of the High-Tc copper oxides. One is based on the t-J Model and its descendants, the other on the Spin-Fermion-Hubbard Model. Most of the researchers apparently chose the former, while our theory is based on the latter. Both possess an RVB-like ground-state, which has been hypothesized in the case of the t-J Model, while in our model it was shown to be an eigenstate of the effective hole Hamiltonian.

The theory we have developed, has provided an accurate description of the many observables that compose the phase diagram for several High-TC cuprate compounds. This also includes a precise description of the effects of an external pressure on the phase diagram. It also provides a unified and universal description of the resistivity and magnetoresistivity of the cuprates in each of the non-superconducting phases [3, 4] of these materials. We derive, for instance,in a very simple and natural way, the intriguing linear-in-$T$ dependence of the resistivity, which is observed in the Strange Metal phase and also predict that the slope is proportional to $T^*$ [3, 4]. Our theory also provides [3, 4] an accurate description of the crossover $H^2$ to $H$, observed in the magnetoresistivity [54] as well as the prediction of the independence of the PG temperature $T^*(x)$ on an applied external pressure [2].

The present work contains, specifically in Sections 2.1, 2.2, 2.3, 3.1 and 3.2, a certain amount of reviews of published material, This has been made for the sake of making it self-contained and easy-to-read. In all of these cases reference to the original publications has been clearly included.

I would like to emphasize, however, the numerous amount of completely new and frequently important results this study contains (mainly in Sections 2.4, 3.3, 3.4, 3.5, 3.6, Section 4 and Appendixes A, B, C): a) the determination of analytic expressions for the transition lines $T_N(x)$, $T_g(x)$ and $T_{CO}(x)$ delimiting the Néel, Spin Glass and charge ordered phases, which together with the previously obtained $T_C(x)$, $T^*(x)$ and $T_{FL}(x)$ form an unprecendetedly complete description of the phase diagram of hole-doped cuprates; b) the obtainment of an explicit expression for a RVB-like state and the demonstration that it is the ground-state of the effective holes' Hamiltonian derived from the Spin-Fermion-Hubbard Model, thus providing the Hamiltonian that completes Anderson's RVB theory; c) The detailed demonstration, in Appendix A, of the stability of the approximation we made by expanding the Hubbard-Stratonovitch fields about their ground-state expectation value; d) the discovery of a complementary relation between the Superconducting and Spin Glass phases in hole doped cuprates; d) the discovery of the mechanism responsible for the onset of a Spin Glass phase in hole-doped cuprates, which allows one to predict that such a phase does not exist in electron doped cuprates, due to the absence of holes.

In our first study about the High-$T_c$ cuprates,[2] when we started to develop our theory for such materials, we made a remark, stating that the study reported there, represented a concrete step forward in the attempt to understand high-Tc superconductivity in cuprates. As we conclude this study, our 6th. publication about these materials, we can assert we have made enough concrete steps so as to reach that goal. We have exhibited enough results to support this assertion. The readers' point-of-view is most welcome.

## A) APPENDIX A: STABILITY ANALYSIS

### A.1) The Stationary Condition

Given a thermodynamic potential such as $\Omega(\Delta, M, \mu)$, we find that the stationary condition corresponds to the three functional derivatives vanishing, as it is expressed by (43).

From expression (32), we then find

$$\frac{\delta \Omega}{\delta |\Delta|} = 2|\Delta| \left\{ \frac{1}{g_S} - f(x_+, x_-) \right\} = 0, \tag{85}$$

$$\frac{\delta \Omega}{\delta |M|} = 2|M| \left\{ \frac{1}{g_P} - g(x_+, x_-) \right\} = 0, \tag{86}$$

and

$$\frac{\delta \Omega}{\delta \mu} = N d(x) - \frac{N}{2T} \left[ \beta_+ \frac{\tanh x_+}{x_+} + \beta_- \frac{\tanh x_-}{x_-} \right], \tag{87}$$

where

$$f(x_+, x_-) = \frac{N}{4T} \left[ \frac{\tanh x_+}{x_+} + \frac{\tanh x_-}{x_-} \right], \tag{88}$$

$$g(x_+, x_-) = \frac{N}{4T} \frac{1}{\sqrt{M^2 + \epsilon^2}} \left[ \beta_+ \frac{\tanh x_+}{x_+} + \beta_- \frac{\tanh x_-}{x_-} \right], \tag{89}$$

$$g(x_+, x_-) = f(x_+, x_-) + \frac{N}{4T} \frac{\mu}{\sqrt{\epsilon^2 + M^2}} \left[ \frac{\tanh x_+}{x_+} - \frac{\tanh x_-}{x_-} \right] \tag{90}$$

where $\beta_\pm = \sqrt{\epsilon^2 + M^2} \pm \mu$ and $x_\pm = \frac{\varepsilon_\pm}{2T}$.

Notice that for $\mu > 0$ we have $\varepsilon_+ > \varepsilon_-$, hence $x_+ > x_-$, whereas for $\mu < 0$ we have $\varepsilon_+ < \varepsilon_-$, hence $x_+ < x_-$. Then, considering that the function

$$\frac{\tanh x}{x}$$

is monotonically decreasing, we find that the second term above is always negative. It follows from (90) the inequality

$$g(x_+, x_-) < f(x_+, x_-) \tag{91}$$

### A.2) THE COMPETITION BETWEEN SC AND PG PHASES

Here we will show that we cannot have both $\Delta \neq 0$ and $M \neq 0$.

In order to do that, let us assume that this is true, namely, that both $\Delta$ and $M$ are nonzero. Then, it follows from (85) and (86), respectively, that

$$f(x_+, x_-) = 1/g_S$$

and

$$g(x_+, x_-) = 1/g_P$$

.

Now, inserting these results in (91), we would immediately conclude that $g_S < g_P$, which is false for all cuprates studied with our theory [2–6].

We conclude, therefore, that we cannot have coexistence of the SC and PG states in cuprates.

### A.3) The Superconducting Phase: $|\Delta| \neq 0; |M| = 0$

We start by considering the SC phase, where $|\Delta| \neq 0; |M| = 0$. From (85) we conclude that, since in the SC phase $\Delta \neq 0$, we must have

$$\frac{1}{g_S} - f(x_+, x_-) = 0, \tag{92}$$

Eq. (86) is automatically satisfied by $M = 0$.

Now, for the stationary solution of (85), (86) and (87) to be actually a minimum of the grand-canonical potential, being, therefore *stable*, we must show that the Hessian matrix of the grand-canonical potential has only positive eigenvalues. A necessary and sufficient condition for this is that all the minor principal determinants of the Hessian matrix should be positive.

The Hessian matrix of $\Omega$ is given by

$$\mathcal{H}(\Delta, M, \mu) = \begin{pmatrix} \Omega_{MM} & \Omega_{M\Delta} & \Omega_{M\mu} \\ \Omega_{\Delta M} & \Omega_{\Delta\Delta} & \Omega_{\Delta\mu} \\ \Omega_{\mu M} & \Omega_{\mu\Delta} & \Omega_{\mu\mu} \end{pmatrix} \tag{93}$$

The elements of the matrix above are the following.

$$\Omega_{\Delta\Delta} = \frac{\delta^2 \Omega}{\delta |\Delta|^2} = \tag{94}$$

$$2[\tfrac{1}{g_S} - f(x_+, x_-)] - 2\Delta \frac{df(x_+, x_-)}{d|\Delta|} = \frac{N}{2T}\Delta^2 \left[ \frac{\alpha_+}{\varepsilon_+^2} + \frac{\alpha_-}{\varepsilon_-^2} \right]$$

where we used (92). In the above expressions, $\alpha_\pm = \alpha(x_\pm)$ and

$$\alpha(x) = \frac{1}{\cosh x}\left[\frac{\sinh x}{x} - \frac{1}{\cosh x}\right] > 0 \tag{95}$$

This is always positive because the first term between brackets is larger than one, while the second is less. We also have, consequently, $\Omega_{\Delta\Delta} > 0$.

Now, let us consider

$$\Omega_{\mu\mu} = \frac{\delta^2 \Omega}{\delta \mu^2} = \frac{N}{2T}\left[\frac{\beta_+^2}{\varepsilon_+^2}\alpha_+ + \frac{\beta_-^2}{\varepsilon_-^2}\alpha_-\right] \tag{96}$$

$$\Omega_{\Delta\mu} = \frac{\delta^2 \Omega}{\delta|\Delta|\delta\mu} = \frac{N}{2T}\Delta\left[\frac{\beta_+}{\varepsilon_+^2}\alpha_+ - \frac{\beta_-}{\varepsilon_-^2}\alpha_-\right] \tag{97}$$

In the above expressions, $\alpha_\pm = \alpha(x_\pm)$ and

$$\alpha(x) = \frac{1}{\cosh x}\left[\frac{\sinh x}{x} - \frac{1}{\cosh x}\right] > 0 \tag{98}$$

This is always positive because the first term between brackets is larger than one, while the second is less.

Then, we have

$$\Omega_{MM} = \frac{\delta^2 \Omega}{\delta |M|^2} = \frac{1}{g_P} - g(x_+, x_-) \quad (99)$$

Using the fact that $g(x_+, x_-) < f(x_+, x_-)$, hence, according to (99) and (92)

$$\Omega_{MM} > \frac{1}{g_P} - f(x_+, x_-) = \frac{1}{g_P} - \frac{1}{g_S} > 0 \quad (100)$$

The last result above follows from the fact that, for all cuprates studied in [2–4], we have $g_S > g_P$.

Using the fact that $M = 0$ in the SC phase, we obtain

$$\Omega_{\Delta M} = \frac{\delta^2 \Omega}{\delta |\Delta| \delta |M|} = \Omega_{M\mu} = \frac{\delta^2 \Omega}{\delta |M| \delta \mu} = 0 \quad (101)$$

We conclude that the Hessian matrix turns out to be block-diagonal, namely,

$$\mathcal{H}(M, \Delta, \mu) = \begin{pmatrix} \Omega_{MM} & 0 & 0 \\ 0 & \Omega_{\Delta\Delta} & \Omega_{\Delta\mu} \\ 0 & \Omega_{\mu\Delta} & \Omega_{\mu\mu} \end{pmatrix} \quad (102)$$
$$= \Omega_{MM} \bigoplus \begin{pmatrix} \Omega_{\Delta\Delta} & \Omega_{\Delta\mu} \\ \Omega_{\mu\Delta} & \Omega_{\mu\mu} \end{pmatrix}$$

The three minor principal determinants of the Hessian matrix are

$$\mathcal{D}_1 = \Omega_{MM}$$

$$\mathcal{D}'_1 = \Omega_{\Delta\Delta}$$

and

$$\mathcal{D}'_2 = [\Omega_{\Delta\Delta}\Omega_{\mu\mu} - \Omega_{\Delta\mu}\Omega_{\mu\Delta}] =$$
$$\left(\frac{N}{2T}\right)^2 \Delta^2 \frac{(\beta_+ \beta_-)^2}{\varepsilon_+^2 \varepsilon_-^2} \alpha_+ \alpha_- \quad (103)$$

Observe that all the minor principals of the Hessian matrix are positive: $\mathcal{D}_1 > 0, \mathcal{D}_2 > 0, \mathcal{D}_3 > 0$ These results guarantee that all eigenvalues of the Hessian matrix are positive, thus implying that the stationary solution we have used for the SC phase is stable, hence our model for the cuprates predicts the existence of a stable SC thermodynamic phase.

### A.4) The Pseudogap Phase: $|\Delta| = 0; |M| \neq 0$

The previous result demonstrated the stability of the SC phase. Let us consider now the PG phase, where $|\Delta| = 0; |M| \neq 0$. For $|\Delta| = 0$, it follows that $\beta_\pm = \varepsilon_\pm$ and also

$$\Omega_{\Delta M} = \frac{\delta^2 \Omega}{\delta |\Delta| \delta |M|} = \Omega_{\Delta\mu} = \frac{\delta^2 \Omega}{\delta |\Delta| \delta \mu} = 0 \quad (104)$$

The Hessian matrix now is given by

$$\mathcal{H}(\Delta, M, \mu) = \begin{pmatrix} \Omega_{\Delta\Delta} & 0 & 0 \\ 0 & \Omega_{MM} & \Omega_{M\mu} \\ 0 & \Omega_{\mu M} & \Omega_{\mu\mu} \end{pmatrix} \quad (105)$$
$$= \Omega_{\Delta\Delta} \bigoplus \begin{pmatrix} \Omega_{MM} & \Omega_{M\mu} \\ \Omega_{\mu M} & \Omega_{\mu\mu} \end{pmatrix}$$

Here

$$\Omega_{\Delta\Delta} = \frac{\delta^2 \Omega}{\delta |\Delta|^2} = \frac{1}{g_S} - f(x_+, x_-) \quad (106)$$
$$\Omega_{\Delta\Delta} < \frac{1}{g_S} - g(x_+, x_-) = \frac{1}{g_S} - \frac{1}{g_P} < 0,$$

where we used (91).

$$\Omega_{MM} = \frac{\delta^2 \Omega}{\delta |M|^2} = \frac{N}{2T} \frac{M^2}{\epsilon^2 + M^2} [\alpha_+ + \alpha_-] \quad (107)$$

$$\Omega_{\mu\mu} = \frac{\delta^2 \Omega}{\delta \mu^2} = \frac{N}{2T} [\alpha_+ + \alpha_-] \quad (108)$$

$$\Omega_{M\mu} = \frac{\delta^2 \Omega}{\delta |M| \delta \mu} = \frac{N}{2T} \frac{M}{\sqrt{\epsilon^2 + M^2}} [\alpha_+ - \alpha_-] \quad (109)$$

The three minor principal determinants of the Hessian matrix are now given by

$$\mathcal{D}_1 = \Omega_{\Delta\Delta}$$

$$\mathcal{D}_2 = \Omega_{MM}$$

$$\mathcal{D}_3 = \Omega_{MM}\Omega_{\mu\mu} - \Omega_{M\mu}\Omega_{\mu M}$$

which is given by

$$[\Omega_{MM}\Omega_{\mu\mu} - \Omega_{M\mu}\Omega_{\mu M}] =$$
$$\left(\frac{N}{2T}\right)^2 \frac{M^2}{\epsilon^2 + M^2} \frac{(\beta_+ \beta_-)^2}{\varepsilon_+^2 \varepsilon_-^2} \alpha_+ \alpha_- \quad (110)$$

We see that the Hessian matrix of the thermodynamic potential $\Omega[\Delta, M, \mu]$ has negative eigenvalues along the direction $\Delta$, whenever $M \neq 0$. This would characterize an instability along this direction. The imposition of the condition $\Delta = 0, M \neq 0, \mu \neq 0$, however, restricts the physical region of the configuration space to the subspace $\Omega[\Delta = 0, M, \mu]$ where the eigenvalues of the sub-Hessian matrix are positive, thus guaranteeing a stable PG state.

## B) APPENDIX B: SECOND ORDER PERTURBATION THEORY IN $J_K$

The Hamiltonian corresponding to the partition functional (11) can be expressed as

$$H = H_{AF} + H_I \tag{111}$$

where

$$H_{AF} = -J_{AF} s^2 \sum_{IJ} \mathbf{n_I}^\alpha \mathbf{n_J}^\alpha \tag{112}$$

and

$$H_I = -J_K s \sum_I \sum_{i \in I} \mathbf{n_I}^\alpha (\mathcal{S}_A - \mathcal{S}_B)_i^\alpha. \tag{113}$$

We are going to perform 2nd. order perturbation theory in $H_I$, in order to obtain the ground state eigenvalue of the Hamiltonian (111), namely

$$E_0 = E_0^{(0)} + E_0^{(1)} + E_0^{(2)} \tag{114}$$

Writing

$$\mathbf{n_I}^\alpha \mathbf{n_J}^\alpha = \frac{1}{2}\left[(\mathbf{n_I} + \mathbf{n_J})^2 - \mathbf{n_I}^2 - \mathbf{n_I}^2\right] \tag{115}$$

we find

$$E_0^{(0)} - E_1^{(0)} = -\frac{J_{AF}}{4}$$

Now using the set of states:

$$|j0\rangle \otimes |j_A m\rangle_A \otimes |j_B m'\rangle_B,$$

namely,

$$\begin{aligned} &|00\rangle \otimes |00\rangle_A \otimes |00\rangle_B \\ &|10\rangle \otimes |1m\rangle_A \otimes |00\rangle_B \\ &|10\rangle \otimes |00\rangle_A \otimes |1m'\rangle_B \end{aligned} \tag{116}$$

where $j = 0, 1$ and $m, m' = 0, \pm 1$, then we will have, according to whether the hole goes into the $A$ or $B$ sublattices

$$\begin{aligned} E_{0A}^{(1)} &=_A \langle 00|\mathcal{S}_A|00\rangle_A = 0 \\ E_{0B}^{(1)} &=_B \langle 00|\mathcal{S}_B|00\rangle_B = 0 \end{aligned} \tag{117}$$

where we used the fact that the $\mathcal{S}_{x,y}$ can be expressed in terms of the raising/lowering operators $\mathcal{S}_+$ and $\mathcal{S}_-$.

$$\begin{aligned} \mathcal{S}_x &= \frac{1}{2}\left(\mathcal{S}_+ + \mathcal{S}_-\right) \\ \mathcal{S}_y &= \frac{1}{2i}\left(\mathcal{S}_+ - \mathcal{S}_-\right) \end{aligned} \tag{118}$$

Now, turning to the 2nd. order correction, we have the following expression for it

$$E_0^{(2)} = \sum_m \frac{\langle 1m_i|H_I|m\rangle\langle m|H_I|1m_f\rangle}{E_0^{(0)} - E_m^{(0)}} \tag{119}$$

It follows that the 2nd. order correction to $J_{AF}$ will be

$$H_{AF}^{(2)} = -J^{(2)} s^2 \sum_{IJ} \mathbf{n_I}^\alpha \mathbf{n_J}^\alpha \tag{120}$$

where $J^{(2)}$ are the eigenvalues of the matrix,

$$E_0 = \pm L_0 \begin{pmatrix} -1 & 0 & 1 \\ 0 & 0 & 0 \\ 1 & 0 & -1 \end{pmatrix} \quad ; \quad L_0 = \frac{4J_K^2}{J_{AF}} \tag{121}$$

where the two overall signs correspond to whether the hole goes into the $A$ or $B$ sublattices. This matrix has eigenvalues: $J^{(2)} = \{0, \pm L_0\}$, hence, after including the 2nd. order correction, the modifications of the energy corresponding to $H_{AF}$ can be reduced to the replacement

$$J_{AF} \longrightarrow \{J_{AF}, J_{AF} + L_0, J_{AF} - L_0\} \tag{122}$$

## C) APPENDIX C: YBCO, A CASE STUDY

YBCO is a singular example of high-Tc cuprate, because of the peculiar characteristics of the doping process. Indeed, up to a stoichiometric doping about $x \simeq 0.115$ the doping process will generate holes into the $CuO_2$ planes. Approximately at this level of doping the holes cease to go into these planes and, instead, all the holes go into chains, which are located out of the $CuO_2$ planes, thereby interrupting the process of $CuO_2$ planes doping. This property effectively makes the YBCO samples to behave in such a way as if it consisted of two different materials, above and below $x \simeq 0.115$.

Entering in expression (44) the data corresponding to the maximum temperature and optimal doping, namely, $T_{max} = 92.9K$, $x_0 = 0.167$, we obtain the green curve in Fig. 21 for $\gamma = 0.076\ eV$. The blue curve in Fig. 21 has been obtained by using (44) for $T_{max} = 61.2K$ and $x_0 = 0.11$ and $\gamma = 0.076\ eV$. Finally, the gold curve in Fig. 21 is obtained by using (44), with $T_{max} = 92.9K$, $x_0 = 0.18$ and $\gamma = 0.041\ eV$.

Let us turn now to the pseudogap (PG) transition temperature $T^*(x)$. We have derived in [2] a universal expression for this temperature, which is given by (47)

For $\tilde{\gamma} = 0.105$ and $\tilde{x}_0 = 0.2213$, we obtain the curve on Fig. 21.

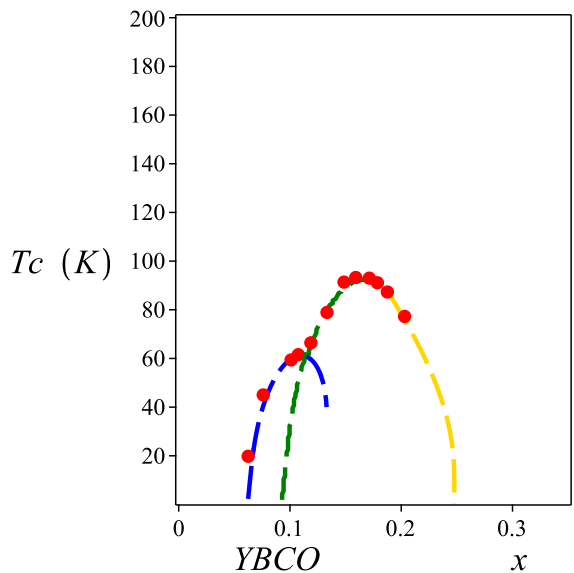

FIG. 21: The YBCO SC dome: $x \in [0, 0.115]$ (blue); $x \in [0.115, 0.19]$ (green); $x \in [0.19, 0.35]$ (gold). Experimental data from [43]

## ACKNOWLEGEMENTS

This work was partially supported by CNPq, FAPERJ and CAPES, brazilian foundations

## COMPETING INTERESTS

The author declares that he has no competing financial interests or personal relationships that could somehow appear to influence the research reported in this article.

---

* marino@if.ufrj.br
[1] J. G. Bednorz and K. A. Müller, *Possible high-Tc superconductivity in the Ba-La-Cu-O system*, Z. Phys. B64, 189 (1986).
[2] E. C. Marino, R. O. Corrêa Jr, R. Arouca, L. H. C. M. Nunes, and V. S. Alves, *Superconducting and Pseudogap Transition Temperatures in High-Tc Cuprates and the Tc Dependence on Pressure*, Supercond. Sci. and Tech. 33, 035009 (2020)
[3] R. Arouca and E. C. Marino *The resistivity of High-Tc Cuprates*, Supercond. Sci. and Tech. 34, 035004 (2021)
[4] E. C. Marino and R. Arouca, *Magnetic Field Effect on the transport properties of high-Tc Cuprates*, Supercond. Sci. and Tech. 34, 085008 (2021)
[5] E. C. Marino *Three Studies in High-Tc Cuprates* New Journal of Physics, 24, 063009 (2022)
[6] E. C. Marino *A Testable Model for High-Tc Superconductivity in Cuprates* SciPost Phys. Proc. 11, 004 (2023) [SCES 2022 Amsterdam (Invited Talk)]
[7] P. W. Anderson, *Resonating valence bonds: A new kind of insulator?*, Materials Research Bulletin 8, 153 (1973); P. Fazekas and P. W. Anderson, *On the ground state properties of the anisotropic triangular antiferromagnet*, Philosophical Magazine 30, 423 (1974).
[8] P. W. Anderson, *The resonating valence bond state in $La_2CuO_4$ and superconductivity*, Science 235, 1196 (1987).
[9] B. Edegger, V. N. Muthukumar and C. Gros, *Gutzwiller-RVB theory of high-temperature superconductivity: Results from renormalized mean-field theory and variational Monte Carlo calculations*, Advances in Physics 56, 927 (2007).
[10] G. Baskaran, Z. Zou and P. W. Anderson, *The resonating-valence-bond state and high-Tc superconductivity* Solid State Commun. 63, 973 (1987)
[11] P. W. Anderson, G. Baskaran, Z. Zou, and T. Hsu, *Resonating-valence-bond theory of phase transitions and superconductivity in $La_2CuO_4$-based compounds* Phys. Rev. Lett. 58, 2790 (1987)
[12] D. S. Rokhsar and S. A. Kivelson, *Superconductivity and the quantum-hard core dimer gas* Phys. Rev. Lett. 61, 2376 (1988)
[13] S. A. Kivelson, D. S. Rokhsar, and J. P. Sethna, *Topology of the resonating-valence-bond ground state: Solitons and high Tc superconductivity.* Phys. Rev. B35, 8865 (1987)
[14] P. W. Anderson, P. A. Lee, M. Randeria, T. M. Rice, N. Trivedi and F. C. Zhang, *The Physics Behind High-Temperature Superconducting Cuprates: The "Plain Vanilla" Version Of RVB* J Phys. Condens. Matter 16 (2004) R755
[15] V.J. Emery, Phys. Rev. Lett. 58, 3759 (1987).
[16] C.M. Varma, S. Schmitt-Rink and E. Abrahams, Solid State Commun. 62, 681 (1987).
[17] N. P. Armitage, P. Fournier, and R. L. Greene, *Progress and perspectives on electron-doped cuprates*, Rev. Mod. Phys. 82, 2421 (2010)
[18] E. C. Marino, *A model for doping in high-temperature superconductors* Phys. Lett. A263, 446 (1999)
[19] E. C. Marino and M. B. S. Neto, *Quantum skyrmions and the destruction of long-range antiferromagnetic order in the high-Tc superconductors LSCO and YBCO*, Phys. Rev. B64, 092511 (2001).
[20] E. C. Marino and L. H. C. M. Nunes *Quantum criticality and superconductivity in quasi-two-dimensional Dirac electronic systems* Nucl. Phys. B741, 404 (2006)
[21] E. C. Marino and L. H. C. M. Nunes *Magnetic field effects on the superconducting and quantum critical properties of layered systems with Dirac electrons* Nucl. Phys. B769, 275 (2007)
[22] L. H. C. M. Nunes, R. L. S. Farias and E. C. Marino, *Superconducting and excitonic quantum phase transitions in doped Dirac electronic systems* Phys. Lett. A376, 779 (2012)
[23] E. C. Marino and L. H. C. M. Nunes *Competing effective interactions of Dirac electrons in the Spin–Fermion system* Annals of Physics 340, 13 (2014)
[24] J. R. Schrieffer and P. A. Wolff, *Relation between the Anderson and Kondo Hamiltonians*, Phys. Rev. 149, 491 (1966)
[25] Sergey Bravyi, David DiVincenzo and Daniel Loss, *Schrieffer-Wolff transformation for quantum many-body systems* Ann. of Phys. 326, 2793 (2011)
[26] K. A. Chao, J. Spalek and A. M. Olés, Phys. Rev. B18,

3453 (1978) *Canonical perturbation expansion of the Hubbard model*

[27] F. C. Zhang and T. M. Rice, *Effective Hamiltonian for the superconducting Cu oxides*, Phys. Rev. B37, 3759 (1988).

[28] L. L. Foldy and S. A. Wouthuysen, *On the Dirac Theory of Spin 1/2 Particles and Its Non- Relativistic Limit*, Phys. Rev. 78, 29 (1950)

[29] V. J. Emery, *Electron doping tests theories* Nature 337, 306 (1989)

[30] J. Zaanen and A. M. Olés, *Canonical perturbation theory and the two-band model for high-Tc, superconductors*, Phys. Rev. B37, 9423 (1988)

[31] J. M. Tranquada, S. M. Heald and A. R. Moodenbaugh, *X-ray-absorption near-edge-structure study of $La_{2x}(Ba,Sr)_xCuO_{4y}$ superconductors* Phys. Rev. B36, 5263 (1987)

[32] J. M. Tranquada, S. M. Heald, A. R. Moodenbaugh, G. Liang and M. Croft, *Nature of the charge carriers in electron-doped copper oxide superconductors* Nature 337, 720 (1989)

[33] A. Aharony, R. J. Birgeneau, A. Coniglio, M. A. Kastner, and H. E. Stanley, *Magnetic phases and magnetic pairing in doped $La_2CuO_4$* Phys. Rev. Lett. 60, 1330 (1988)

[34] W. Su, J. R. Schrieffer and A. J. Heeger, *Solitons in polyacetylene*, Phys. Rev. Lett. 42, 1698 (1979).

[35] A. H. Castro Neto and D. Hone, *Doped Planar Quantum Antiferromagnets with Striped Phases* Phys. Rev. Lett. 76, 2165 (1996)

[36] N. D. Mermin and H. Wagner *Absence of Ferromagnetism or Antiferromagnetism in One- or Two-Dimensional Isotropic Heisenberg Models*, Phys. Rev. Lett., 17, 1133 (1966)

[37] M. K. Chan, C. J. Dorow, L. Mangin-Thro, Y. Tang, Y. Ge, M. J. Veit, G. Yu, X. Zhao, A. D. Christianson, J. T. Park, Y. Sidis, P. Steffens, D. L. Abernathy, P. Bourges and M. Greven *Commensurate antiferromagnetic excitations as a signature of the pseudogap in the tetragonal high-Tc cuprate $HgBa_2CuO_{4+\delta}$* Nature Communications 7, 10819 (2016)

[38] E. Stilp, A. Suter, T. Prokscha, E. Morenzoni, H. Keller, B. M. Wojek, H. Luetkens, A. Gozar, G. Logvenov, and I. Bozovic, *Magnetic phase diagram of low-doped $La_{2x}Sr_xCuO_4$ thin films studied by low-energy muon-spin rotation*, Phys. Rev. B88, 064419 (2013)

[39] R. B. Laughlin, *Anomalous Quantum Hall Effect: An Incompressible Quantum Fluid with Fractionally Charged Excitations*, Phys. Rev. Lett. 50, 1395

[40] E.C Marino, *Quantum Field Theory Approach to Condensed Matter Physics*, Cambridge University Press, Cambridge, UK (2017)

[41] M. Hucker, V. Kataev, J. Pommer, U. Ammerhal, A. Revcolevschi, J. M. Tranquada, and B. Buchner, Phys. Rev. B70, 214515 (2004)

[42] Ch. Niedermayer, C. Bernhard, T. Blasius, A. Golnik, A. Moodenbaugh, and J. I. Budnick, *Common Phase Diagram for Antiferromagnetism in LSCO and YBCO as Seen by Muon Spin Rotation*, Phys. Rev. Lett. 80, 3843 (1998)

[43] O. Cyr-Choinière, R. Daou, F. Laliberté, C. Collignon, S. Badoux, D. LeBoeuf, J. Chang, B. J. Ramshaw, D. A. Bonn, W. N. Hardy, R. Liang, J.-Q. Yan, J.-G. Cheng, J.-S. Zhou, J. B. Goodenough, S. Pyon, T. Takayama, H. Takagi, N. Doiron-Leyraud, and Louis Taillefer, *Pseudogap temperature $T^*$ of cuprate superconductors from the Nernst effect* Phys. Rev. B97, 064502 (2018)

[44] J.-J. Wen, H. Huang, S.-J. Lee, H.Jang, J. Knight, Y.S. Lee, M. Fujita, K.M. Suzuki, S. Asano, S.A. Kivelson, C.-C. Kao1 and J.-S. Lee, *Observation of two types of charge-density-wave orders in superconducting $La_{2-x}Sr_xCuO_4$* Nature Communications 10, 3269 (2019)

[45] H. Murayama, Y. Sato, R. Kurihara, S. Kasahara, Y. Mizukami, Y. Kasahara, H. Uchiyama, A. Yamamoto, E.-G. Moon, J. Cai, J. Freyermuth, M. Greven, T. Shibauchi and Y. Matsuda *Diagonal nematicity in the pseudogap phase of $HgBa_2CuO_{4+\delta}$*, Nature Communications 10, 3282 (2019)

[46] C.M.S. da Conceição and E.C. Marino, *Stable mean-field solution of a short-range interacting SO (3) quantum Heisenberg spin glass* Phys. Rev. Lett. 101, 037201 (2010) .

[47] C.M.S. da Conceição and E.C. Marino, *Duality, quantum skyrmions, and the stability of a SO (3) two-dimensional quantum spin glass* Phys. Rev. B80, 064422 (2009)

[48] C.M.S. da Conceição and E.C. Marino, *Quantum field theory solution for a short-range interacting SO(3) quantum spin-glass* Nucl. Phys. B82, 565-592 (2009)

[49] K. von Arx, Q. Wang, S. Mustafi, D. G. Mazzone, M. Horio, D. Mukkattukavil, E. Pomjakushina, S. Pyon, T. Takayama, H. Takagi, T. Kurosawa, N. Momono, M. Oda, N. B. Brookes, D. Betto, W. Zhang, T. C. Asmara, Y. Tseng, T. Schmitt, Y. Sassa and J. Chang, *Fate of charge order in overdoped La-based cuprates*, NPJ Quantum Materials 8, 7 2023

[50] W. Tabis, B. Yu, I. Bialo, M. Bluschke, T. Kolodziej, A. Kozlowski, E. Blackburn, K. Sen, E. M. Forgan, M. v. Zimmermann, Y. Tang, E. Weschke, B. Vignolle, M. Hepting, H. Gretarsson, R. Sutarto, F. He, M. Le Tacon, N. Barisic, G. Yu, and M. Greven *Synchrotron x-ray scattering study of charge-density-wave order in $HgBa_2CuO_{4+\delta}$* Phys. Rev. B96, 134510 (2017)

[51] W. Tabis, Y. Li, M. Le Tacon, L. Braicovich, A. Kreyssig, M. Minola, G. Dellea, E. Weschke, M.J. Veit, M. Ramazanoglu, A.I. Goldman, T. Schmitt, G. Ghiringhelli, N. Barisic, M.K. Chan, C.J. Dorow, G. Yu, X. Zhao, B. Keimer and M. Greven *Charge order and its connection with Fermi-liquid charge transport in a pristine high-Tc cuprate* Nature Commun. 5, 5875 (2014)

[52] H. Miao, R. Fumagalli, M. Rossi, J. Lorenzana, G. Seibold, F. Yakhou-Harris, K. Kummer, G. D. Gu, L. Braicovich, G. Ghiringhelli, and M. P. M. Dean, *Formation of Incommensurate Charge Density Waves in Cuprates* Phys. Rev. X9, 031042 (2019)

[53] H. Miao, G. Fabbris, R. J. Koch, D. G. Mazzone, C. S. Nelson, R. Acevedo-Esteves, G. D. Gu, Y. Li1, T. Yilimaz, K. Kaznatcheev, E. Vescovo, M. Oda, T. Kurosawa, N. Momono, T. Assefa, I. K. Robinson, E. S. Bozin, J. M. Tranquada, P. D. Johnson and M. P. M. Dean, *Charge density waves in cuprate superconductors beyond the critical doping* NPJ Quantum Materials 6, 31 (2021)

[54] J. Ayres, M. Berben, M. Culo, Y-T. Hsu, Y-T. van Heumen, Y-T. Huang, J. Zaanen, T. Kondo, T. Takeuchi, C. Cooper JR, S. Putzkel, S. Friedemann, T. Carrington and N. E. Hussey *Incoherent transport across the strange metal regime of highly overdoped cuprates.* Nature 595, 661 (2021).